\pdfoutput=1 
\documentclass[superscriptaddress,showpacs,showkeys,pra,twocolumn,amsmath,amssymb,letterpaper]{revtex4}
\usepackage{times}
\usepackage{hyperref}
\usepackage{soul,xspace,braket}
\usepackage{graphicx}
\usepackage{hypernat}
\usepackage[figure,table]{hypcap}

\hypersetup{colorlinks=true,
linkcolor=blue,
citecolor=blue,
urlcolor=blue,
pdfauthor={Matthias Rosenkranz, Alexander Klein, and Dieter Jaksch},
pdftitle={Simulating and detecting artificial magnetic fields in
  trapped atoms},
pdfkeywords={quantum simulation, polaron, two-species Bose-Hubbard
  model, artificial magnetic fields}}

\mathchardef\ordinarycolon\mathcode`\:
\mathcode`\:=\string"8000
\begingroup \catcode`\:=\active
  \gdef:{\mathrel{\mathop\ordinarycolon}}
\endgroup

\newcommand{\BEC}{\caps{BEC}\xspace}
\newcommand{\GPE}{\caps{GPE}\xspace}
\newcommand{\GOE}{\caps{GOE}\xspace}
\newcommand{\di}{\ensuremath{d}}
\newcommand{\im}{\ensuremath{i}}
\newcommand{\eu}{\ensuremath{\mathrm{e}}}

\newcommand{\bvec}[1]{\ensuremath{\mathbf{#1}}}

\begin{document}
\title{Simulating and detecting artificial magnetic fields in trapped atoms}
\author{Matthias Rosenkranz}
\email{m.rosenkranz@physics.ox.ac.uk}

\affiliation{Clarendon Laboratory, University of Oxford, Parks Road,
  Oxford OX1 3PU, United Kingdom}

\affiliation{Keble College, Parks
  Road, Oxford OX1 3PG, United Kingdom}

\author{Alexander Klein}
\affiliation{Clarendon Laboratory, University of Oxford, Parks Road,
  Oxford OX1 3PU, United Kingdom}

\affiliation{Keble College, Parks
  Road, Oxford OX1 3PG, United Kingdom}

\author{Dieter Jaksch}
\homepage{http://www.physics.ox.ac.uk/qubit/}

\affiliation{Clarendon Laboratory, University of Oxford, Parks Road,
  Oxford OX1 3PU, United Kingdom}

\affiliation{Keble College, Parks
  Road, Oxford OX1 3PG, United Kingdom}

\affiliation{Centre for Quantum Technologies, National University of
  Singapore, 117543, Singapore}



\date{\today}

\begin{abstract}
  A Bose-Einstein condensate exhibiting a nontrivial phase induces an
  artificial magnetic field in immersed impurity atoms trapped in a
  stationary, ring-shaped optical lattice.  We present an effective
  Hamiltonian for the impurities for two condensate setups: the
  condensate in a rotating ring and in an excited rotational state in
  a stationary ring.  We use Bogoliubov theory to derive analytical
  formulas for the induced artificial magnetic field and the hopping
  amplitude in the limit of low condensate temperature where the
  impurity dynamics is coherent.  As methods for observing the
  artificial magnetic field we discuss time of flight imaging and mass
  current measurements.  Moreover, we compare the analytical results
  of the effective model to numerical results of a corresponding
  two-species Bose-Hubbard model.  We also study numerically the
  clustering properties of the impurities and the quantum chaotic
  behavior of the two-species Bose-Hubbard model.
\end{abstract}
\maketitle

\section{Introduction}
A prominent application of ultracold atoms trapped in optical lattices
is that of a quantum simulator of condensed matter models.  The
advantage of such a simulator is that the Hamiltonian defining the
condensed matter model can be implemented almost perfectly, and the
relevant parameters, such as interaction strength or lattice geometry,
can be tuned in a wide parameter
range~\cite{Lewenstein-AiP-2006,Jaksch-Ann-2005}.  In contrast, for a
given condensed matter system the Hamiltonian governing its dynamics
is often difficult to find.  Examples of effects which can be
simulated with neutral atoms in optical lattice are the quantum Hall
effect in two
dimensions~\cite{vonKlitzing-PRL-1980,Sorensen-PRL-2005,Palmer-PRL-2006,Palmer-PRA-2008}
or persistent ground state currents in superconducting
rings~\cite{SchMueUst97, RyuAndCla07}.

Several interesting effects in condensed matter experiments typically
require a system to be exposed to a magnetic field.  Therefore, a
quantum simulator of these effects needs to include or simulate such a
magnetic field.  Ultracold atoms offer several promising ways to
achieve this goal.  One idea is to rotate a gas of atoms close to the
frequency of the harmonic confinement that traps the atoms.  This may
result in states also found in the quantum Hall
effect~\cite{WilGun00,Cooper-PRL-2001,Schweikhard-PRL-2004}.  However,
the high rotation frequency $\Omega$ of the atoms introduces a large
centrifugal term, which tends to destroy the atomic cloud.  On the
other hand, to reach the strongly correlated state the rotation
frequency of the cloud and the frequency of the confining harmonic
trap $\omega_c$ have to be balanced with high accuracy such that
$1-Z/2N\tilde a_s \alt \Omega/\omega_c \alt 1-8\tilde a_s/ZN$, where
$Z$ is the transverse distance over which the condensate is uniform,
$N$ is the number of atoms, and $\tilde a_s$ is on the order of the
s-wave scattering length~\cite{RosPetSin02}.  The necessity of precise
counterbalancing is alleviated, in principle, by using traps with
potentials stronger than harmonic at the expense of harder
experimental implementation~\cite{Bretin-PRL-2004}.  A different
approach is to use lasers with an orbital angular
momentum~\cite{Juzeliunas-PRL-2004, Juzeliunas-PRA-2005} or to use
Raman-assisted hopping in a layered optical lattice
geometry~\cite{Jaksch-NJP-2003} in order to mimic the effect of a
magnetic field.  Both proposals might also be difficult to realize
experimentally.

In this paper, we analyze and extend methods introduced by some of the
present authors~\cite{KleJak09} for simulating a magnetic field in
atoms trapped in a ring-shaped, stationary optical lattice.  These
lattice atoms, in the following also called impurities, are immersed
into a ring-shaped quasi-onedimensional Bose-Einstein condensate (\BEC)
exhibiting a nontrivial phase.  For creating this phase we consider
two possible setups.  Firstly, a BEC trapped in a rotating
ring~\cite{MorColLor06,Gupta-PRL-2005,Heathcote-NJP-2008}; secondly, a
BEC carrying a quantized angular momentum in a stationary
ring~\cite{NanWalSch04}, i.e., a BEC in an excited state.  In the
former case, we can allow a confining potential to overcompensate the
centrifugal term in the rotating BEC as we are mainly interested in
the dynamics of the impurities.  This is in contrast to methods for
observing quantum Hall states directly in a rotating gas of ultracold
atoms where the centrifugal term needs to be counterbalanced
precisely~\cite{RosPetSin02}.  The BEC is described in terms of
Bogoliubov excitations, which couple to the impurity atoms.  It has
been shown in earlier publications that the presence of the
phonon-impurity coupling leads to observable effects in the transport
properties of the lattice atoms such as the transition from coherent
to incoherent transport with increasing temperature or
clustering~\cite{Bruderer-PRA-2007,Bruderer-NJP-2008,Klein-NJP-2007}.
In addition, by allowing the BEC to exhibit a nontrivial phase the
coupling leads to an artificial magnetic field in the dynamics of the
impurities which manifests itself as a nontrivial phase term in the
hopping of the lattice atoms.  By describing the impurities in terms
of polarons we derive an analytical expression for the induced phase.

The induced phase changes the dynamics of the lattice atoms
considerably.  We show that it leads to current carrying ground states
in the impurities, which can be detected experimentally.  Another
experimentally accessible method for detecting the phase of ultracold
atoms is to evaluate time of flight (ToF)
images~\cite{Nunnenkamp-PRA-2008}.  We discuss this method as a means
of revealing the rotational state of the underlying BEC with the
impurity as a probe.  We find that nonzero temperature and increasing
interspecies coupling obscure this transition.  We also show that the
tendency of the lattice atoms to assemble in clusters depends on the
value of the induced phase.  Furthermore, we observe the onset of
quantum chaos if the symmetry of our two-species system is broken by a
disorder potential.  Quantum chaos is characterized by the
distribution of energy levels.  If this distribution is of the
Gaussian orthogonal ensemble (\GOE) type, then the system is said to
behave quantum chaotically~\cite{BohGiaSch84,KolBuc04}.  Although we
restrict ourselves to a quasi-onedimensional system throughout this
paper, our onedimensional (1D) model will allow insights into
experimentally more realistic twodimensional (2D) setups.

This paper is organized as follows.  In Sec.~\ref{sec:model} we derive
an effective Hamiltonian for the impurity atoms immersed into the
\BEC.  The impurities in this model are dressed with a cloud of \BEC
excitations, which induces additional interactions and a phase term on
the impurity hopping.  The dressed impurities can be interpreted as
polarons~\cite{Bruderer-PRA-2007,Mahan-2000}.  We derive analytical
formulas for the induced phase in different experimental setups.  In
Sec.~\ref{sec:probing} we present two methods for detecting the
additional phase twist in the impurities.  These methods are based on
ground state current measurements and ToF imaging.  In
Sec.~\ref{sec:clustering} we study the effect of the phonon-mediated
interaction in the impurities on their clustering properties.
Finally, we extend previous studies~\cite{Buonsante-PRA-2008} of the
chaotic behavior of a one-species Bose-Hubbard model with a phase
twist to two species in Sec.~\ref{sec:chaos}.  We conclude in
Sec.~\ref{sec:conclusion}.  The application of the Bogoliubov
approximation to the Bose-Hubbard model with a phase twist is
discussed in an Appendix.

\section{The model}\label{sec:model}
We consider two species of ultracold atoms, species $a$ and $b$, both
confined in a quasi-1D ring geometry.  Atoms of species $b$ form a
\BEC confined in a ring trap.  Atoms of species $a$, considered as
impurities in the \BEC, are trapped in a ring-shaped optical
lattice~\cite{FraLeaPad07}.  This optical lattice is submerged into
the \BEC which is not affected by the presence of the optical lattice
potential~\cite{Leblanc-PRA-2007}.  For a detailed derivation of the
model in higher dimensions we refer to
Refs.~\cite{KleJak09,Bruderer-PRA-2007,Bruderer-NJP-2008}.  The term
quasi-1D is taken to mean that the atoms are confined in a tight
transverse trap such that they are in the ground state of this trap.
A transverse harmonic trap with frequency $\omega_\perp$ introduces
the characteristic length scale $\ell_\perp =
\sqrt{\hbar/m_b\omega_\perp}$, where $m_b$ is the mass of atoms of
species $b$.  For the gas to be in the transverse ground state of this
trap we require that $\hbar\omega_\perp \gg gn$, where $g =
2\hbar^2a_s/m_b\ell_\perp^2$ is the 1D interaction strength with $a_s$
the three-dimensional (3D) s-wave scattering length and $n=N/L$ the
linear density of the gas in a ring of length $L$ with $N$ atoms.
Here we have assumed that the scattering of the atoms is still
essentially 3D, i.e., $\ell_\perp \gg a_s$.  In lower dimensional
gases phase and density fluctuations play an important role.  Our
calculations are valid if the phase of the \BEC does not fluctuate
over the circumference of the ring $L$.  In a 1D \BEC coherence is
preserved over a length scale $\ell_c = n\hbar^2/m_bk_BT$, where $k_B$
is Boltzmann's constant and $T$ is the \BEC
temperature~\cite{PetShlWal00}.  As we will employ the Bogoliubov
approximation, density fluctuations must be small.  This is ensured in
the weakly interacting limit $n\xi_h \gg 1$, with $\xi_h =
\hbar/\sqrt{m_bgn}$ the healing length, if $T \ll T_d =
\sqrt{\hbar^2n^3g/m_b}/k_B$~\cite{Cas04}.  In this temperature regime
the atoms are in a \BEC because $T_d$ lies below the condensation
temperature~\cite{BayBlaHol99,ZobRos06}.

We write the Hamiltonian of the full system as $\hat H = \hat H_B +
\hat H_a + \hat H_I$, where the three parts represent the \BEC, the
impurities, and the interaction between the two species, respectively.
The \BEC Hamiltonian $\hat H_B$ and interaction $\hat H_I$ are given
by
\begin{align*}
  \hat H_B &= \int\di x \hat\phi^\dagger(x) \left[\hat H_0 +
    \frac{g}{2} \hat\phi^\dagger(x) \hat\phi(x) \right] \hat\phi(x),\\
  \hat H_I &= \kappa\int\di x \hat\chi^\dagger(x)\hat\chi(x)
  \hat\phi^\dagger(x)\hat\phi(x).
\end{align*}
The field operator $\hat\phi(x)$ annihilates a \BEC atom at position
$x$, whereas $\hat\chi(x)$ annihilates an impurity at $x$.  The
Hamiltonian $\hat H_0$ describes the noninteracting part of the \BEC
and will be defined when we discuss specific examples, $g$ is the 1D
interaction strength of the \BEC atoms and $\kappa$ is the 1D
interaction strength between a \BEC atom and an impurity.  We assume
that the temperature is sufficiently low such that the impurities in
the lattice only occupy the lowest band.  Then we can expand their
field operator in terms of Wannier functions $\eta_j(x)$
as~\cite{Koh59}
\begin{equation*}
  \hat\chi(x) = \sum_j \eta_j(x) \hat a_j.
\end{equation*}
The operator $\hat a_j$ annihilates an impurity in lattice site
$j$. We assume the lattice to be sufficiently deep such that the
tight-binding approximation for the Wannier functions holds, i.e.,
$\int\di x |\eta_j(x)|^2|\eta_{j'}(x)|^2 \simeq 0$ for different
lattice sites $j \neq j'$.  With these assumptions the impurities are
well described by a Bose-Hubbard Hamiltonian~\cite{Jaksch-PRL-1998}
\begin{equation*}
  \hat H_a = -\tilde J_a\sum_{\langle j,j'\rangle} \hat a_j^\dagger
  \hat a_{j'}^{\vphantom{\dagger}} + \frac{\tilde U_a}{2} \sum_j \hat
  n_j (\hat n_j-1) - \tilde \mu_a \sum_j \hat n_j,
\end{equation*}
where $\tilde J_a$ is the hopping matrix element between two
neighboring sites, $\tilde U_a$ the onsite interaction, $\hat n_j =
\hat a_j^\dagger \hat a_j^{\vphantom{\dagger}}$, and $\tilde\mu_a$ is
the chemical potential.  We note that it is straightforward to extend
our derivation to fermionic impurities by using a Fermi-Hubbard
Hamiltonian.

We write the \BEC field operator as $\hat\phi(x) = \phi_0(x) +
\hat\zeta(x)$, where $\phi_0(x)$ is the solution of the
Gross-Pitaevskii equation (\GPE) for $\kappa=0$
\begin{equation}\label{eq:GPE}
  [\hat H_0 + g|\phi_0(x)|^2] \phi_0(x) = 0
\end{equation}
and $\hat\zeta(x)$ describes a small quantized deviation from the mean
field solution $\phi_0(x)$.  We will be studying rotating \BEC{}s,
which means that $\phi_0(x)$ is, in general, a complex function.
Furthermore, we assume that the impurity-boson coupling $\kappa$
fulfills $|\kappa|/gn_0(x) \xi_h(x) \ll 1$, where $n_0(x) =
|\phi_0(x)|^2$ is the mean field density and we allow a
space-dependent density in the healing length.  The smallness
condition for the interspecies coupling $\kappa$ ensures that the
deformation of the \BEC due to the presence of the impurities is
small~\cite{Bruderer-NJP-2008}.  We plug the expansion $\phi_0(x) +
\hat\zeta(x)$ into the Hamiltonian $\hat H_B + \hat H_I$ and keep
terms up to second order in $\kappa$.  The linear terms in
$\hat\zeta(x)$ result in the interspecies interaction $\kappa\int\di x
\hat\chi^\dagger(x)\hat\chi(x) [\phi_0(x) \hat\zeta^\dagger(x) +
\phi_0^*(x) \hat\zeta(x)]$.  The quadratic terms in $\hat\zeta(x)$ are
diagonalized by expanding the fluctuations in terms of Bogoliubov
excitations as $\hat\zeta(x) = \sideset{}{'}\sum_q[u_q(x) \hat b_q -
v_q^*(x) \hat b_q^\dagger]$. Here $\hat b_q$ annihilates an excitation
(phonon) in mode $q$ and the prime indicates that the sum does not
include the \BEC mode.  The operators $\hat b_q$ and $\hat
b_q^\dagger$ fulfill the usual bosonic commutation relations $[\hat
b_q^{\phantom{\dagger}}, \hat b_{q'}^\dagger] = \delta_{qq'}$ and
$[\hat b_q, \hat b_{q'}] = 0$.  The mode functions $u_q(x)$ and
$v_q(x)$ satisfy the Bogoliubov-de Gennes (BdG)
equations~\cite{KleJak09}
\begin{subequations}\label{eq:BdG}
\begin{align}
  [\hat H_0 + 2gn_0(x)] u_q(x) - g[\phi_0(x)]^2 v_q(x) &=
  \hbar\omega_q u_q(x),\\
  [\hat H_0^\dagger + 2gn_0(x)] v_q(x) - g[\phi_0^*(x)]^2 u_q(x) &=
  -\hbar\omega_q v_q(x).
\end{align}
\end{subequations}
The eigenenergies of these equations are the quasiparticle energies
$\hbar\omega_q$.  The total Hamiltonian is then given by
the Hubbard-Holstein Hamiltonian~\cite{Holstein-Ann-1959}
\begin{equation}\label{eq:H-HH}
\begin{split}
  \hat H &= \hat H_a + \sum_j\sideset{}{'}\sum_q \hbar\omega_q
  (M_{j,q} \hat b_q + M_{j,q}^* \hat b_q^\dagger) \hat n_j\\
  &\quad + \sum_j \bar E_j \hat n_j + \sideset{}{'}\sum_q
  \hbar\omega_q^{\phantom{\dagger}} \hat b_q^\dagger \hat
  b_q^{\phantom{\dagger}}.
\end{split}
\end{equation}
Here, $M_{j,q} = (\kappa/\hbar\omega_q) \int \di x [\phi_0^*(x) u_q(x)
- \phi_0(x) v_q(x)] |\eta_j(x)|^2$ are matrix elements of the
phonon-impurity coupling, and $\bar E_j = \kappa\int \di x n_0(x)
|\eta_j(x)|^2$ is a mean field shift.  By using the Lang-Firsov
transformation~\cite{Mahan-2000} this Hamiltonian can be shown to
describe the dynamics of polarons in an optical
lattice~\cite{KleJak09,Bruderer-PRA-2007}.  The polarons in this model
are given by the impurity atoms surrounded by a coherent cloud of
Bogoliubov phonons.  In the following we require that $k_B T \ll E_j$
and $\tilde J_a \ll E_j$, where $E_j= E_p = \sideset{}{'}\sum_q
\hbar\omega_q |M_{j,q}|^2$ is the polaronic level shift.  For the
cases considered in this work it is independent of the lattice site
$j$.  We assume the phonons to be thermally occupied at temperature
$T$.  Furthermore, the characteristic hopping speed has to fulfill
$d\tilde J_a/\hbar \ll c$, where $c \sim \sqrt{gn_0/m_b}$ is the speed
of sound and $d$ the lattice spacing.  These conditions ensure that
the polaron dynamics is coherent, retardation effects are suppressed,
and that the hopping term can be treated as a
perturbation~\cite{Bruderer-PRA-2007}.  As shown in
Ref.~\cite{KleJak09} this allows us to write down the effective
Hamiltonian
\begin{equation}\label{eq:H_eff}
\begin{split}
  \hat H_\text{eff} &= -J_a\sum_{\langle j,j'\rangle} \eu^{2\pi\im
    \alpha_{j,j'}}
  \hat a_j^\dagger \hat a_{j'}^{\phantom{\dagger}} + \frac{1}{2}\sum_j U_j \hat n_j (\hat n_j-1)\\
  &\quad - \sum_j \mu_j \hat n_j - \frac{1}{2} \sum_{j\neq j'}
  V_{j,j'} \hat n_j \hat n_{j'}.
\end{split}
\end{equation}
The chemical potential is $\mu_j = \tilde\mu_a - \kappa n_0 + E_j$,
$U_j = \tilde U_a - 2E_j$, the reduced hopping $J_a = \tilde J_a
\exp\left( -\sideset{}{'}\sum_q |M_{j,q} - M_{j',q}|^2 [2N_q(T) + 1]/2
\right)$, where $N_q(T) = [\exp(\hbar\omega_q/k_B T)-1]^{-1}$, and the
interaction $V_{j,j'} = \sideset{}{'}\sum_q \hbar\omega_q
(M_{j,q}M_{j',q}^* + M_{j,q}^*M_{j',q})$.  The Boltzmann distribution
$N_q(T)$ enters because we have averaged the \BEC degrees of freedom
over a thermal phonon distribution at temperature $T$.  The phase
factor in the hopping term of Eq.~\eqref{eq:H_eff} is given by
\begin{equation}\label{eq:alpha_ij}
  \alpha_{j,j'} = \frac{1}{4\pi\im}
  \sideset{}{'}\sum_q (M_{j,q}^{\vphantom{*}}M_{j',q}^* -
  M_{j,q}^*M_{j',q}^{\vphantom{*}}).
\end{equation}
Since we allow the condensate wave function $\phi_0(x)$ to be complex,
in general, $\alpha_{j,j'}$ is nonzero.  This means that the impurity
atoms pick up a phase when they hop across lattice sites.  The
derivation above demonstrates that the origin of this induced phase is
a coupling of quasiparticles in the \BEC to the impurities.  The
occurrence of such a phase in the Bose-Hubbard model leads to a change
in the dynamics of the impurities in the lattice.  In the following we
will consider concrete systems and derive the analytical expressions
for the corresponding induced phases.

\subsection{The BEC in a rotating ring}\label{ssec:ring-model}
For the first model system we restate here the results
of~\cite{KleJak09} for a \BEC in a rotating ring and also introduce
quantities necessary for the discussion in the remainder of the paper.
The quasi-1D \BEC is confined to a ring of length $L$ rotating at
angular frequency $\Omega$. The impurities are trapped in a ring-shaped
optical lattice, which is submerged into the \BEC.  The noninteracting
part of the Hamiltonian of the \BEC is
\begin{equation}\label{eq:H_0-rot}
  \hat H_0 = -\frac{\hbar^2}{2m_b} \frac{\di ^2}{\di x^2} +
  \im\hbar v \frac{\di}{\di x} - \mu_b^\text{rot},
\end{equation}
where $v = R\Omega$ is the rotational speed, $R = L/2\pi$ is the
radius of the ring, and $\mu_b^\text{rot}$ the chemical potential of
the \BEC.  The function $\phi_0(x) = \sqrt{n_0} \exp(\im k x)$ solves
the corresponding \GPE, Eq.~\eqref{eq:GPE}, with the quantized
momentum $k = 2\pi m/L$ and $m$ an integer such that $k = m_bv/\hbar -
\Delta k$, where $\Delta k \in [-\pi/L, \pi/L)$.  The density of the
\BEC is denoted with $n_0$.  This definition ensures that the \BEC is
in the ground state with its momentum closest to the momentum of the
rotation $m_bv/\hbar$.  The chemical potential of the \BEC is given by
$\mu_b^\text{rot} = \hbar^2k^2/2m_b + gn_0 - \hbar vk$.  The
fluctuations around the ground state wave function lead to Bogoliubov
excitations with energy
\begin{equation}\label{eq:homega-rotating}
  \hbar\omega_q^\text{rot} = E_q^B - \frac{\hbar^2q\Delta k}{m_b},
\end{equation}
where $E_q^B = \sqrt{\epsilon_q^0(\epsilon_q^0 + 2gn_0)}$ and
$\epsilon_q^0 = \hbar^2q^2/2m_b$.  This Bogoliubov phonon energy
contains the mismatch $\Delta k$ between the ground state momentum and
the angular momentum given by the rotation speed.  We rewrite the
definition of the mismatch as $\Delta v := \hbar\Delta k/m_b = v -
\hbar k/m_b =: v - v_b$, where $v_b$ is the velocity of the \BEC.
Positive excitation energies require that $|\Delta v| \leq \min_{q\neq
  0} E_q^B/\hbar q \sim c$.  This asymptotic behavior of the velocity
follows if we only consider low lying excitations (since $L \gg
\xi_h$).  The Bogoliubov modes satisfy the periodic boundary
conditions, which results in the quantization of their quasimomenta as
$q = 2\pi m/L$ with $m$ an integer.  The solutions of the BdG
equations, Eqs.~\eqref{eq:BdG}, are the mode functions
\begin{subequations}\label{eq:modes-rot}
\begin{align}
  u_q(x) &= \frac{u_q}{\sqrt{L}} \eu^{\im (q+k)x},\\
  v_q(x) &= \frac{v_k}{\sqrt{L}} \eu^{\im (q-k)x},
\end{align}
\end{subequations}
where $u_q^2 = 1+v_q^2 = [(\epsilon_q^0+gn_0)/E_q^B + 1]/2$.  Their
coefficients fulfill $u_q \pm v_q = (E_q^B/\epsilon_q^0)^{\pm 1/2}$.
In a tightly confining lattice the Wannier functions of the impurities
are well described by Gaussians $\eta_j(x) =
\exp[-(x-x_j)^2/2\sigma^2]/\sqrt{\sqrt{\pi}\sigma}$ with width
$\sigma$ centered at a lattice site $j$.  The position of the $j$-th
lattice site is parametrized by its angle $\Phi_j$ on the ring, i.e.,
$x_j = \Phi_j N_s d/2\pi$, where $N_s$ is the number of lattice sites
and $d$ is the lattice spacing.  The coupling matrix elements are then
given by
\begin{equation*}
  M_{j,q}^\text{rot} = \frac{\kappa}{\hbar\omega_q^\text{rot}}
  \sqrt{\frac{n_0}{L}}\sqrt{\frac{\epsilon_q^0}{E_q^B}} \eu^{\im qx_j - q^2\sigma^2/4}.
\end{equation*}
The Gaussian width $\sigma$ of the localized impurities introduces a
cutoff in the phonon momenta contributing to the coupling.  By
plugging this result into Eq.~\eqref{eq:alpha_ij} we finally arrive at
the induced phase of the impurities
\begin{equation}\label{eq:alpha-rotating}
  \alpha_a^\text{rot} = \frac{\kappa^2 n_0}{2\pi L} \sideset{}{'}\sum_q
  \frac{\epsilon_q^0}{E_q^B} \frac{1}{(\hbar\omega_q^\text{rot})^2}
  \eu^{-q^2\sigma^2/2} \sin(qd),
\end{equation}
where $\alpha_a^\text{rot} = \alpha_{j+1,j}$ since we only
consider nearest-neighbor hopping.  As a result of the momentum cutoff
in the coupling terms, the momenta contributing to the sum in the
phase are distributed with a Gaussian envelope.  The reduced hopping
term is given by
\begin{equation}\label{eq:J_a-rot}
\begin{split}
  J_a^\text{rot} &= \tilde J_a\exp\Biggl(-\frac{\kappa^2 n_0}{L}
  \sideset{}{'}\sum_q \frac{\epsilon_q^0}{E_q^B}
  \frac{1}{(\hbar\omega_q^\text{rot})^2} \eu^{-q^2\sigma^2/2}\\
  &\quad \times[1-\cos(qd)] [2N_q(T) + 1] \Biggr).
\end{split}
\end{equation}

In Fig.~\ref{fig:alpha_L}(a) we plot the induced phase for typical
experimental parameters for different \BEC coupling strengths
$g$~\cite{MadCheBre01,HodHecHop01}.  The values of the induced phase
$\alpha_a^\text{rot} \simeq 0.03$ are sufficiently large to cause
observable effects in the dynamics of the system.  In our coherent
approximation, the effect of nonzero temperature is only to reduce the
effective hopping rate.  For simplicity we have set the temperature to
zero in the figure and the magnitude of the hopping $J_a^\text{rot}$
does not deviate significantly from the magnitude of the bare hopping
$\tilde J_a$.  Incoherent hopping will only affect the phase if the
temperature lies above the polaron energy~\cite{Bruderer-PRA-2007}.
Density fluctuations will play a significant role for temperatures $T
> T_d$.  If the length of the ring exceeds the coherence length
$\ell_c$, the assumption of a well defined phase over the whole \BEC
is invalidated.  The temperature scale for thermal phonon excitation
is of similar magnitude.  Furthermore, temperatures at $k_BT \geq
n\pi\hbar^2/m_bR$ will excite additional rotations of the \BEC, which
results in a multi-peaked distribution of rotation velocities and a
gradual loss of superfluidity~\cite{CarCas04}. Typically, all these
effects start playing a significant role at temperatures on the order
of a few tens of nanokelvin.  Hence, the effects discussed here can be
observed for sufficiently cold samples.

From Fig.~\ref{fig:alpha_L}(a) we see that at first the induced phase
depends linearly on the rotation frequency of the \BEC.  At the
critical frequency $\Omega_\text{crit} = \hbar/2m_bR^2$ we observe a
sudden jump to negative phases.  This jump is caused by a jump in the
momentum of the \BEC at the critical rotation frequency.  Because of
the quantization of the quasimomentum in units of $2\pi/L$, at this
frequency the ground state of the \BEC changes to a nonzero
quasimomentum.  These jumps repeat at odd multiples of the critical
rotation frequency.

\subsection{The BEC in an optical lattice}\label{ssec:BEC-lattice}
\begin{figure}
  \centering%
  \includegraphics[width=\linewidth]{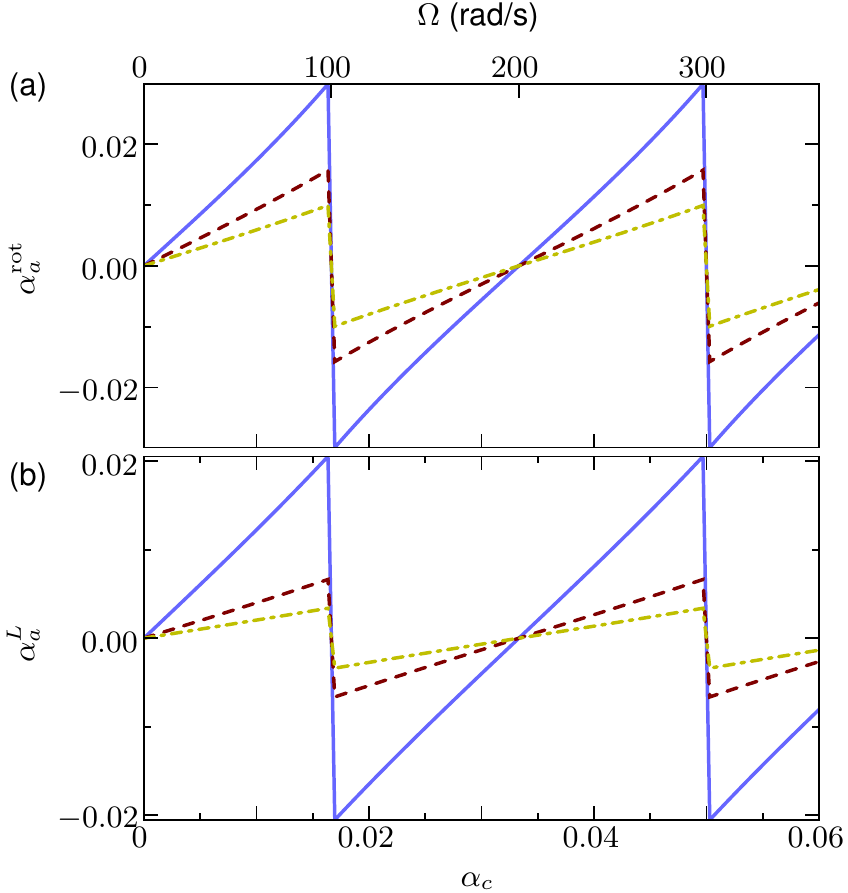}
  \caption{(Color online) Induced phase in the lattice with $30$
    lattice sites for (a) a ring rotating with angular speed $\Omega$
    and (b) the \BEC in a lattice with phase term $\alpha_c$.  We
    assume a \BEC of $^{87}$Rb with a linear density $n_0 = 5\times
    10^6~\mathrm{m}^{-1}$.  The impurities are $^{87}$Rb atoms in a
    different hyperfine level.  The lattice spacing is $d =
    400~\mathrm{nm}$, which means that the ring has a circumference of
    $L = 12~\mathrm{\mu m}$.  In (a) the interspecies coupling
    strength is $\kappa/2dE_R = 0.045$, where $E_R =
    \hbar^2\pi^2/2m_ad^2$ is the recoil energy of the impurity lattice
    with impurity mass $m_a$, and the curves indicate different \BEC
    couplings $g/2dE_R = 0.01, 0.015, 0.02$ (solid, dashed,
    dashed-dotted lines, respectively). In (b) we assume that both
    lattices have the same depth, interspecies coupling is $U_I/J_c =
    1.5$, and \BEC couplings $U_c/J_c = 0.5, 1, 1.5$ (solid, dashed,
    and dashed-dotted lines, respectively).}\label{fig:alpha_L}
\end{figure}

In order to be able to compare our analytical results with numerical
solutions we now introduce a system which is numerically solvable
beyond the Bogoliubov approximation.  This allows us to study the
validity of our model of an induced phase beyond the Bogoliubov
approximation.  We assume that the \BEC consisting of $N_c$ atoms is
trapped in a ring-shaped optical lattice with $N_s$ lattice sites and
spacing $d$, which is independent of the lattice for the impurities.
We will derive an analytical expression for the induced phase in the
Bogoliubov approximation, which will have a similar form as the
expression in the continuous case of Sec.~\ref{ssec:ring-model},
Eq.~\eqref{eq:alpha-rotating}.  We will use these results for the
discussion of the numerical results in Sec.~\ref{sec:probing}.  One
can show that in the limit of $N_s\rightarrow\infty$ with $N_sd = L =
\mathrm{const}$ the two expressions coincide.

The Hamiltonian of the condensate is given by a generalized
Bose-Hubbard Hamiltonian~\cite{Rey-PRA-2007}
\begin{equation}\label{eq:H_BH}
\begin{split}
  \hat H_{BH} &= -J_c \sum_j \left( \eu^{2\pi\im\alpha_c} \hat
    c_{j+1}^\dagger\hat c_j^{\phantom{\dagger}} +
    \eu^{-2\pi\im\alpha_c} \hat c_j^\dagger\hat c_{j+1}^{\phantom{\dagger}} \right)\\
  &\quad + \frac{U_c}{2} \sum_j^{\phantom{\dagger}} \hat c_j^\dagger
  \hat c_j^{\phantom{\dagger}}(\hat c_j^\dagger \hat
  c_j^{\phantom{\dagger}} - 1),
\end{split}
\end{equation}
where $J_c$ and $U_c$ are the hopping and onsite interaction terms of
the \BEC, respectively, and $\hat c_j$ is the bosonic annihilation
operator for site $j$.  For a lattice rotating with angular frequency
$\Omega$ the phase factor $\alpha_c$ is given by $\alpha_c = m_b
\Omega N_sd^2/4\pi^2 \hbar$~\cite{Rey-PRA-2007}.  The impurities are
trapped in a second lattice as before, for which we assume that it has
the same spacing $d$ and the same number of lattice sites $N_s$ as the
\BEC lattice.  In this picture the interaction between impurity and
\BEC atoms is given by
\begin{equation}\label{eq:H_I}
  \hat H_I = U_I \sum_j \hat n_j \hat c_j^\dagger \hat c_j,
\end{equation}
where $U_I$ is the coupling strength between the two species.  In the
remainder of this section we will derive analytically the induced
phase of this system resulting from our approximate polaron model.
Later we will use Eqs.~\eqref{eq:H_BH} and \eqref{eq:H_I} for
numerical computations of the full two-species model.

Similar to the implementation discussed in the preceding section we
now expand the annihilation operators $\hat c_j$ around an order
parameter $\phi_j$ as $\hat c_j = \phi_j + \hat \xi_j$. Then we
rewrite this expansion in the momentum representation with
quasimomentum $2\pi q/N_s$, where $q$ is an integer, and apply a
Bogoliubov transformation with quasiparticle operators $\hat b_q$ and
$\hat b_q^\dagger$.  For the fluctuations to be small we require that
$U_I/U_c \bar n_0 \ll 1$, where $\bar n_0 = N_c/N_s$ is the number of
\BEC atoms per lattice site.  The main result is that, as before, we
arrive at an effective Hamiltonian of the Hubbard-Holstein form with
polarons as the quasiparticles hopping across lattice sites.  A
derivation of the results is given in the Appendix.  In the coherent
dynamics limit the Bogoliubov excitations mediate additional
interactions and induce a phase in the impurities.  The excitations in
this system have an energy
\begin{equation}\label{eq:homega_L}
  \hbar\omega_q^L = E_q^L + \Lambda_q,
\end{equation}
where $E_q^L = \sqrt{\epsilon_q^L (\epsilon_q^L + 2 U_c \bar n_0)}$,
$\Lambda_q = -2J_c \sin(2\pi q/N_s) \sin(\Delta\theta/N_s)$, and
$\epsilon_q^L = 4J_c\sin^2(\pi q/N_s)\cos(\Delta\theta/N_s)$.  The
angle $\Delta\theta$ is determined via the relation $2\pi\alpha_c =
2\pi \nu/N_s + \Delta\theta/N_s$ with $\nu$ an integer and
$\Delta\theta \in [-\pi, \pi)$.  The ground state momentum is
determined by the winding number $\nu = \lfloor\alpha_c N_s\rceil$,
where the symbol $\lfloor\cdot\rceil$ denotes rounding to the nearest
integer.  One can show that the phonon energies in
Eq.~\eqref{eq:homega_L}, reduce to the excitation energies of the
rotating ring, Eq.~\eqref{eq:homega-rotating}, in the limit
$N_s\rightarrow\infty$ with $N_sd = L = \text{const}$.  The coupling
matrix elements between impurities and phonons are given by
\begin{equation}\label{eq:M-lattice}
  M_{j,q}^L = \frac{U_I}{\hbar\omega_q^L} \sqrt{\frac{\bar n_0}{N_s}}
  \sqrt{\frac{\epsilon_q^L}{E_q^L}} \eu^{2\pi\im q j/N_s}.
\end{equation}
From Eq.~\eqref{eq:alpha_ij} we calculate the induced phase of the
impurity atoms as
\begin{equation}\label{eq:alpha-lattice}
  \alpha_a^L = \frac{U_I^2 \bar n_0}{2\pi N_s} \sum_{q=1}^{N_s-1}
  \frac{\epsilon_q^L}{E_q^L} \frac{1}{(\hbar\omega_q^L)^2} 
  \sin(2\pi q/N_s).
\end{equation}
The form of this induced phase is very similar to the one for a \BEC
in a rotating ring, Eq.~\eqref{eq:alpha-rotating}.  Here however, the
cutoff in the phonon quasimomentum is introduced by the quantization
of the momentum through the finite lattice.  This results in a finite
sum instead of a Gaussian depending on the impurity localization width
$\sigma$.  The reduced hopping term is given by
\begin{equation*}
\begin{split}
  J_a^L &= \tilde J_a \exp\Biggl( - \frac{U_I^2 \bar n_0}{N_s}
  \sum_{q=1}^{N_s-1} \frac{\epsilon_q^L}{E_q^L}
  \frac{1}{(\hbar\omega_q^L)^2}\\
  &\quad \times [1 - \cos(2\pi q/N_s)] [2N_q(T) + 1] \Biggr).
\end{split}
\end{equation*}

Numerical tests show that the rescaled induced phase $\alpha_a^L 2\pi
N_s/U_I^2\bar n_0$ becomes maximal for $\Delta\theta \rightarrow \pi$
and $U_c\bar n_0 \rightarrow 0$.  However, Bogoliubov theory loses
validity at $\Delta\theta = \pi$ because the ground state becomes
degenerate.  At small $U_c/J_c$ we find $\bar n_0 \simeq \bar n =
N_c/N_s$ so in order to maximize the constant factor in the induce
phase, we have to choose a large filling $\bar n$ and large
interspecies coupling $U_I$.  However, the interspecies coupling has
to remain sufficiently small to fulfill the condition $U_I/U_c\bar n
\ll 1$ that was necessary for the analytical derivation.

In Fig.~\ref{fig:alpha_L}(b) we plot the induced phase according to
Eq.~\eqref{eq:alpha-lattice} for similar parameters as in
Fig.~\ref{fig:alpha_L}(a) but now with a \BEC in a rotating optical
lattice.  The induced phase now critically depends on the \BEC phase
$\alpha_c$, which takes the role of the frequency in the preceding
section.  We observe jumps in the induced phase at critical \BEC
phases, which have the same origin as the critical frequencies in the
rotating ring.  The critical phases occur at $\alpha_\text{crit} =
(2j+1)/2N_s$, where $j$ is an integer.  We can expect similar orders
of magnitude for the induced phase for both setups with or without an
optical lattice for the \BEC atoms.  Note that both the upper and
lower x-axes in the figure are valid for both plots because we assume
two species of the same mass and the phase in the Bose-Hubbard
Hamiltonian $\hat H_{BH}$ is a linear function of the frequency.

\subsection{The rotating BEC in a static ring}
\begin{figure}
  \centering
  \includegraphics[width=\linewidth]{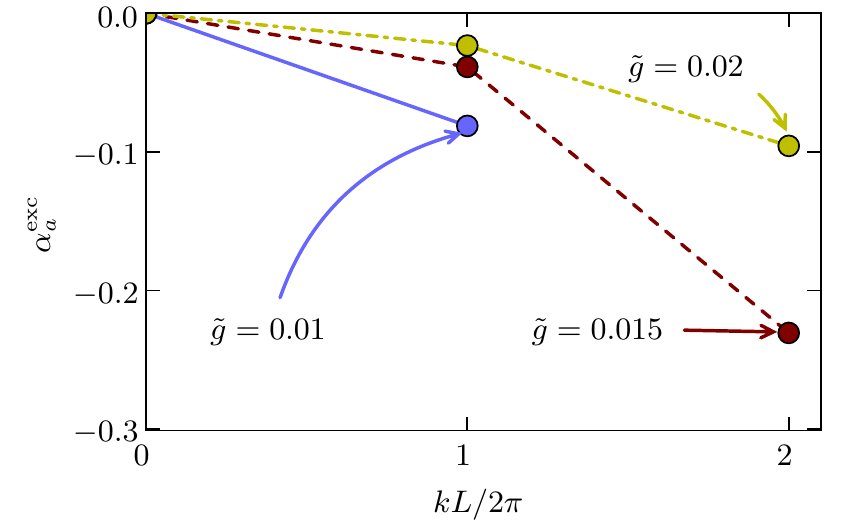}
  \caption{(Color online) Induced phase with a \BEC in an excited
    state.  The parameters are the same as in
    Fig.~\ref{fig:alpha_L}(a) and we abbreviate $\tilde g =
    g/2dE_R$. The critical momenta for the different interaction
    strengths are given by $k_\text{crit}L/2\pi \simeq 2.1$, $2.6$,
    $3.0$ (for $\tilde g = 0.01, 0.015, 0.02$, respectively).  The
    lines are only to guide the eye.}\label{fig:alpha_a_exc}
\end{figure}

Again we consider a quasi-1D \BEC in a ring of radius $R = L/2\pi$ and
the impurities trapped in a ring-shaped optical lattice immersed into
the \BEC.  In contrast to Sec.~\ref{ssec:ring-model}, here we assume
that the \BEC rotates but the ring does not.  Hence, the \BEC is not
in the ground state of the Hamiltonian.  This excited state can be
created, for example, with a \caps{STIRAP}
process~\cite{NanWalSch04,Juzeliunas-PRA-2005} or by initializing the
\BEC in a rotating ring whose rotation is then turned off for the
duration of the experiment.  The interaction-free Hamiltonian has the
same form as the one in the rotating ring, Eq.~\eqref{eq:H_0-rot}, but
with $v=0$.  As before we expand the fluctuations around the order
parameter in terms of Bogoliubov excitations.  The \BEC field operator
is then given by
\begin{equation}\label{eq:phi-exc}
  \hat\phi(x) = \eu^{\im kx} \left[\sqrt{n_0} + \sideset{}{'}\sum_q \left(u_q(x)
      \hat b_q - v_q^*(x) \hat b_q^\dagger\right) \right].
\end{equation}
Periodicity requires the quantization of the phase $k = 2\pi m/L$ with
$m$ an integer. Plugging this ansatz into the \GPE results in the
chemical potential $\mu_b^\text{exc} = \hbar^2k^2/2m_b + gn_0$.  The
solutions of the BdG equations diagonalize the quadratic part of the
Hamiltonian.  In \cite{Fetter-AP-1972} they are given as
\begin{align*}
  u_q(x) &= \frac{u_q}{\sqrt{L}} \eu^{\im qx},\\
  v_q(x) &= \frac{v_q}{\sqrt{L}} \eu^{\im qx}
\end{align*}
with the same coefficients $u_q$, $v_q$ as in the case of a rotating
ring [see definition below Eqs.~\eqref{eq:modes-rot}].  The energies
of these solutions are given by
\begin{equation*}
  \hbar\omega_q^\text{exc} = E_q^B + \hbar qw,
\end{equation*}
where $w = \hbar k/m_b$ is the speed associated with momentum $k$.
For $\hbar\omega_q^\text{exc}$ to be positive we have assumed that the
velocity of the \BEC is less than the speed of sound, i.e., $w < c$,
which is equivalent to a critical momentum $k_\text{crit} = 1/\xi_h$.
Again we assume that the impurity wave function is given by a Gaussian
centered at the position $x_j$ of lattice site $j$.  This allows us to
evaluate the induced phase as
\begin{equation*}
  \alpha_a^\text{exc} = \frac{\kappa^2 n_0}{2\pi L} \sideset{}{'}\sum_q
  \frac{\epsilon_q^0}{E_q^B} \frac{1}{(\hbar\omega_q^\text{exc})^2}
  \eu^{-q^2\sigma^2/2} \sin(qd).
\end{equation*}
We see that the form of this phase is the same as in
Eq.~\eqref{eq:alpha-rotating} but now the phonon energy depends on the
full momentum of the \BEC $k$, not only the mismatch to the ground
state $\Delta k$.  Similarly, the form of the reduced hopping is the
same as Eq.~\eqref{eq:J_a-rot} but with $\hbar\omega_q^\text{rot}$
replaced by $\hbar\omega_q^\text{exc}$.  The higher energies allow for
larger induced phases but entail the experimental difficulty of
maintaining the \BEC in an excited state~\cite{OegKav09,RyuAndCla07}.
A derivation of the life time of this excited state is beyond the
scope of the present work.

Figure~\ref{fig:alpha_a_exc} shows the induced phase in this setup for
experimentally accessible parameters below the critical momentum.
Induced phases up to $\alpha_a^\text{exc} \simeq -0.25$ are readily
achievable in this setup and even higher phase are possible by
slightly changing the parameters.  The reason for the negative sign of
the phase is that the dispersion favors phonons with negative
quasimomentum.  A negative quasimomentum, i.e., total phonon
quasimomentum $k-|q|$, pushes the system closer to the ground state at
zero total momentum.

\section{Probing the phase twist}\label{sec:probing}
In \cite{KleJak09} it was shown that the presence of a phase twist in
the impurity Hamiltonian can be observed as a drift of the impurities.
If we prepare a Gaussian-shaped distribution of impurity atoms
centered at lattice site $j$, then it will expand and the mean
position of this packet will drift either to the left or right
depending on the induced phase.  This drift can be verified
experimentally.  It is also present for impurity speeds below the
speed of sound, which, at first glance, seems to contradict the Landau
criterion of superfluid flow~\cite{Lan41,Leg99}.  From this criterion
one would expect that a \BEC rotating below the Landau critical speed
does not impart a momentum on the impurity.  Recent work by Sykes et
al.~\cite{SykDavRob09} and earlier works by Roberts et
al.~\cite{RobPom05,Rob06} suggest Doppler-shifted scattering processes
of quantum fluctuations in the \BEC with the impurity as a source for
this drift.  When an incoming wave of a \BEC quantum fluctuation is
reflected off an impurity it experiences a Doppler shift depending on
the propagation direction, which leads to an overall drag force on the
impurity.  In contrast to their derivation, our results are based on
the nonperturbative Lang-Firsov transformation and we assume periodic
boundary conditions.

In the remainder of this section, we present detection methods for the
induced phase twist, namely, via mass current measurements and ToF
expansion.  In order to compare the effective one-species model,
Eq.~\eqref{eq:H_eff}, with numerical results we use a two-species
Bose-Hubbard Hamiltonian $\hat H_{2BH} = \hat H_a + \hat H_{BH}$,
which can be solved by numerical diagonalization for small systems.

\subsection{Impurity ground  state momentum}\label{ssec:current}
For a single impurity our effective one-species model predicts a
vanishing ground state momentum for $|\alpha_a| < 1/2N_s$.  In the
full two-species model interactions between the two atom species cause
a broadening of the reduced impurity ground state in momentum space
even for a single impurity.  We now investigate how this broadening
can be used to measure the induced phase $\alpha_a$ of our effective
model.  For vanishing induced phase the broadening in momentum space
is symmetric around the zero momentum state.  However, for $0 <
|\alpha_a| < 1/2N_s$ the broadening becomes asymmetric around the zero
momentum state.  This leads to a small nonvanishing mean momentum of
the impurity ground state.  We note that the origin of this effect is
different from the expected jump of the momentum ground state at the
critical phases $\alpha_\text{crit}$.  At a critical phase the
macroscopic occupation of a momentum state changes, whereas the
asymmetric broadening is a perturbative effect and does not cause such
a macroscopic shift.  To see this effect analytically we calculate the
ground state of the Hubbard-Holstein Hamiltonian, Eq.~\eqref{eq:H-HH},
in first-order Rayleigh-Schr\" odinger perturbation theory for a
single impurity~\cite{KupWhi63}.  For simplicity we let $\bar E_j = 0$
and $\tilde\mu_a = 0$ so that we can rewrite Eq.~\eqref{eq:H-HH} in
momentum space as $\hat H_\text{HH} = \sum_{k} E_k^a \hat a_k^\dagger
\hat a_{k}^{\vphantom{\dagger}} +
\sideset{}{'}\sum_q\hbar\omega_q^\text{rot} \hat b_q^\dagger \hat
b_q^{\vphantom{\dagger}} + \sideset{}{'}\sum_{q,k}
\hbar\omega_q^\text{rot} M_q^\text{rot} \left(\hat
  b_q^{\vphantom{\dagger}} \hat a_{k+q}^\dagger \hat
  a_k^{\vphantom{\dagger}} + \hat b_q^\dagger \hat a_{k-q}^\dagger
  \hat a_k^{\vphantom{\dagger}} \right)$, where $\hat a_k$ annihilates
an impurity with quasimomentum $k$, $E_k^a = -2\tilde J_a \cos(kd)$ is
the unperturbed impurity dispersion with quasimomentum $k$, and
$M_q^\text{rot}$ is the Fourier-transformed coupling matrix element
$M_{j,q}$.  The coupling term mixes phonon and impurity excitations.
We write the unperturbed states as $\ket{k,N_q}$, which indicates an
impurity with quasimomentum $k$ and $N_q$ Bogoliubov excitations with
quasimomentum $q$.  With this notation the perturbed (dressed) ground
state is
\begin{equation*}
  \ket{k_0,0}^{(1)} = \ket{k_0,0} + \sideset{}{'}\sum_q
  \frac{\hbar\omega_q M_q}{E^a_{k_0} - E^a_{k_0-q} - \hbar\omega_q}
  \ket{k_0-q,1_q},
\end{equation*}
where $k_0$ is the quasimomentum of the impurity ground state.  The
quasimomentum distribution $n_k$ and mean quasimomentum $\bar k$ of
this state are given by
\begin{align}
  n_{k} &= \langle \hat a_k^\dagger \hat a_k\rangle = \delta_{kk_0} +
  \left(\frac{\hbar\omega_{k_0-k}|M_{k_0-k}|}{E^a_{k_0}
      - E^a_k - \hbar\omega_{k_0-k}} \right)^2,\label{eq:n_k}\\
  \bar k &= \sum_{k} n_{k} k.\label{eq:bar-k}
\end{align}
To be able to compare the numerically obtained mean quasimomentum with
this formula we insert the expression for $M_k$ for the \BEC in an
optical lattice, Eq.~\eqref{eq:M-lattice}, into Eq.~\eqref{eq:n_k}.
Moreover, in a lattice the quasimomentum $k$ is expressed as $2\pi
k/N_s$, where $k$ is now an integer.  With the resulting quasimomentum
distribution $n_k$ the mean quasimomentum, Eq.~\eqref{eq:bar-k}, reads
\begin{equation}\label{eq:mean_k}
  \bar k = -\frac{U_I^2 \bar n_0}{N_s} \sideset{}{'}\sum_{k=-\lfloor
    N_s/2\rceil+1}^{\lfloor N_s/2\rceil-1}
  \frac{\epsilon_{k}^L}{E_{k}^L} \frac{2\pi k/N_s}{\left(E^a_0 - E^a_{k} -
      \hbar\omega_{k}^L \right)^2},
\end{equation}
where we have used the fact that $\epsilon_{k}^L$, $E_{k}^L$, and
$E_k^a$ are even functions of $k$.  This expression should be compared
to the expression for the induced phase, Eq.~\eqref{eq:alpha-lattice}.
The sine in Eq.~\eqref{eq:alpha-lattice} can be linearized for $2\pi
q/N_s$ close the roots of the sine.  Furthermore, the summation
indices of both expressions can be shifted so that the two expressions
formally coincide apart from the different energy denominators and a
constant.  To analyze the energy denominator in Eq.~\eqref{eq:mean_k}
we assume $U_c = 0$ for clarity.  Then it is given by $E^a_0 - E^a_{k}
- \hbar\omega^L_{k} = -2(\tilde J_a+J_c)[1-\cos(2\pi k/N_s)] -
\Lambda_{k} \simeq -\hbar\omega^L_{k}(1+\tilde J_a/J_c)$ for phases
$\alpha_c$ close to $j/N_s$ ($j$ integer), that is small
$\Lambda_{k}$.  We might thus expect that
\begin{equation}\label{eq:bar-k-prop}
  \bar k \simeq -\frac{2\pi}{(1 + \tilde J_a/J_c)^2} \alpha_a^L.
\end{equation}

\begin{figure}
\includegraphics[width=\linewidth]{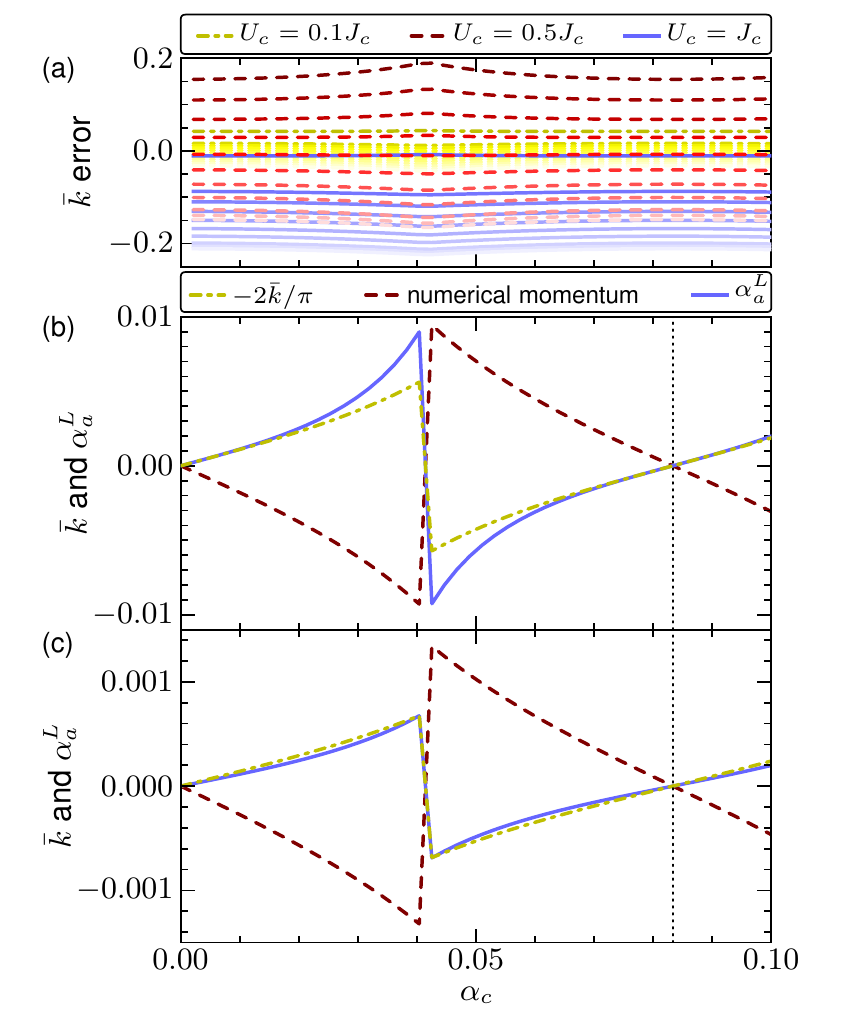}
\caption{(Color online) (a) Relative error $\bar k_\text{ana}/\bar
  k_\text{num} - 1$ of Eq.~\eqref{eq:mean_k} compared to the
  numerically exact ground state quasimomentum for different
  interactions at $N_s = 12$, $6$ \BEC atoms, and one impurity with
  $\tilde J_a = J_c$.  Line styles indicate different \BEC interactions
  $U_c$ (see legend) and increasing lightness (from top to bottom)
  indicates decreasing interspecies coupling $U_I/J_c \in \{0.2, 0.15,
  0.14, \dots, 0.05, 0.01\}$ (dashed-dotted lines), $\{0.9, 0.8,
  \dots, 0.1, 0.05, 0.01\}$ (dashed), $\{1, 0.7, 0.6, \dots, 0.1,
  0.05, 0.01\}$ (solid).  (b)--(c) Comparison of analytical induced
  phase (solid), numerically exact mean quasimomentum (dashed) and the
  rescaled analytical mean quasimomentum $-2\bar k/\pi$
  (dashed-dotted) for (b) $U_c = 0.5J_c$, $U_I = 0.4J_c$ and (c) $U_c
  = J_c$, $U_I = 0.2J_c$.  The rescaling factor $-2/\pi$ results from
  Eq.~\eqref{eq:bar-k-prop}.  The dotted vertical line indicates
  $\alpha_c = 1/N_s$.}\label{fig:momentum}
\end{figure}

In order to check the validity of these approximations we have
computed ground states of the two-species Bose-Hubbard Hamiltonian
with a single impurity.  From the reduced density matrix of the
impurity we compute the mean quasimomentum and compare it with
Eq.~\eqref{eq:mean_k}.  Figure~\ref{fig:momentum}(a) shows the error
of the formula~\eqref{eq:mean_k} relative to the numerical results for
different sets of interactions.  As expected, the accuracy of the
analytical result is better for small interactions (e.g.,
approximately $\pm 5\%$ for $U_c = 0.1J_c$).  In
Fig.~\ref{fig:momentum}(b)--(c) we compare the mean quasimomentum of
the impurity ground state with the analytical result for the induced
phase, Eq.~\eqref{eq:alpha-lattice}.  Clearly, our approximations
discussed in the preceding paragraph predict the value of the induced
phase for $\alpha_a^L$ close to $\alpha_c = j/N_s$ ($j$ integer), that
is away from the critical phases $\alpha_\text{crit} = (2j+1)/2N_s$,
where Bogoliubov theory fails.  Also note that in
Fig.~\ref{fig:momentum}(b) the constraints on the validity of our
analytical derivation of the induced phase are not met ($U_I \not\ll
U_c\bar n_0$).  Nevertheless, the rescaled mean quasimomentum
approximates the induced phase well close to the roots of
$\alpha_a^L$, which means that qualitatively our effective polaron
model still describes the underlying physics.

Our computational resources restrict the size of the systems studied
here to moderately small number of lattice sites and filling factors.
Therefore, we could not directly observe the jump of the impurity to a
macroscopic occupation of a nonzero momentum ground state for
$\alpha_a^L > \alpha_\text{crit}$.  However, our results clearly
indicate an imbalanced momentum distribution owing to the influence of
an induced phase.  For larger number of lattice sites the critical
phase $\alpha_\text{crit}$ decreases.  Furthermore, the quantization
of the impurity ground state momentum in units of $2\pi/N_s$ becomes
denser for larger lattices so that more allowed momentum states become
available for the impurity.  Therefore, we expect to see a macroscopic
population of a nonzero momentum state in larger systems caused by a
large induced phase as predicted by our effective model.  Such a
macroscopic population will be observable as an impurity current in an
experiment.  The current measured by the operator
\begin{equation*}
  \hat C_a = \frac{\im \tilde J_am_a}{\hbar} \sum_{j} \left( \hat a_{j+1}^\dagger
    \hat a_j^{\phantom{\dagger}} - \hat a_j^\dagger \hat a_{j+1}^{\phantom{\dagger}} \right)
\end{equation*}
is closely related to the mean momentum of the impurity.  It is
straightforward to show that the current of a single impurity with
momentum $\bar k$ is $\langle \hat C_a\rangle = 2\tilde J_a
m_a\sin(2\pi\bar k)/\hbar$.  For small $\bar k$ as in
Fig.~\ref{fig:momentum} we may approximate $\langle \hat C_a\rangle
\propto \sin(2\pi\bar k) \simeq 2\pi\bar k$.  In the preceding
paragraph we have seen that close to the zero crossing the induced
phase and mean momentum are proportional, which shows that the induced
phase and the mass current measured by $\hat C_a$ are proportional
around $\alpha_c = 1/N_s$.

\subsection{Time of flight images}\label{ssec:ToF}
Another experimentally readily accessible method to probe for a phase
term is the ToF expansion of the atoms.  After their evolution the
atoms are abruptly released from the trap and expand for a time $t$
before they are imaged.  If interactions can be neglected during the
time of flight, the imaged density distribution corresponds to the
momentum distribution of the atoms in the trap.  This distribution at
a position $\bvec x$ is given by
\begin{equation*}
  \rho_\text{ToF}(\bvec x) = |\tilde w(\bvec K)|^2 \sum_{j_1, j_2}
  \rho_1(\bvec R_{j_1}, \bvec R_{j_2}) \eu^{\im \bvec K\cdot (\bvec R_{j_1} - \bvec    R_{j_2})},
\end{equation*}
where we neglect a constant factor $(m_a/\hbar
t)^3$~\cite{Bloch-RMP-2008}.  In the ballistic approximation the
momentum is $\bvec K = m_a \bvec x/\hbar t$ with $m_a$ the mass of an
impurity atom, and $\tilde w(\bvec K)$ the Fourier transform of the
Wannier function of the impurity trapped in an optical lattice.  The
one-particle density matrix is given by $\rho_1(\bvec R_{j_1}, \bvec
R_{j_2}) = \langle \hat a_{\bvec R_{j_1}}^\dagger \hat a_{\bvec
  R_{j_2}}^{\phantom{\dagger}}\rangle$, where $\hat a^\dagger_{\bvec
  R_j}$ creates an impurity atom at position $\bvec R_j$.  In a ring
of radius $R$ the positions $\bvec R_j$ are fixed at $\bvec R_j = R
[\sin(2\pi j/N_s), \cos(2\pi j/N_s)]$.  For tightly localized atoms we
can set $\tilde w(\bvec K) = 1/\sqrt{N_s}$ and write
\begin{equation*}
\begin{split}
  \rho_\text{ToF}(\bvec x) &= \frac{1}{N_s} \sum_{j_1,j_2}
  \rho_1(\bvec R_{j_1}, \bvec R_{j_2})\\
  &\quad \times \exp\{ \im K_x R [\sin(2\pi j_1/N_s) - \sin(2\pi
  j_2/N_s)]\\
  &\qquad + \im K_y R [\cos(2\pi j_1/N_s) - \cos(2\pi j_2/N_s)] \}.
\end{split}
\end{equation*}

\begin{figure}
  \centering%
  \includegraphics[width=\linewidth]{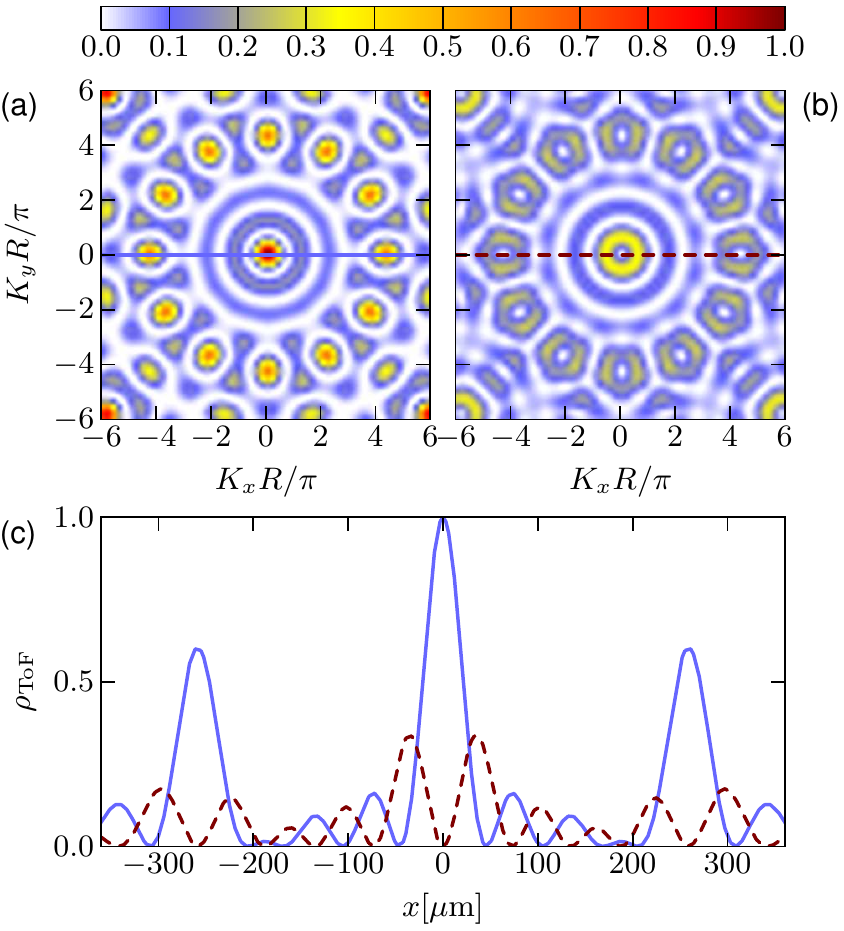}%
  \caption{(Color online) Time of flight distribution
    $\rho_\text{ToF}$ of a single impurity released from a ring
    lattice with $N_s = 12$ at $T=0$ in the single-species model,
    Eq.~\eqref{eq:H_eff}.  The central peak in (a) indicates a zero
    quasimomentum ground state for $\alpha_a = 0.04$.  In (b) the
    vanishing density in the center indicates that the ground state
    exhibits nonzero momentum ($\alpha_a=0.05$).  The jump between
    these two momentum states occurs at $\alpha_a=1/2N_s\simeq 0.042$.
    In (c) we show the corresponding density profiles indicated with
    horizontal lines in (a) and (b).  The ToF expansion time is $t =
    50~\mathrm{ms}$ for $^{87}$Rb atoms in a ring of radius
    $R=12\mu\mathrm{m}$.}\label{fig:ToF}
\end{figure}

To illustrate the effect of an induced phase term in the effective
impurity Hamiltonian, Eq.~\eqref{eq:H_eff}, we first consider the ToF
expansion of a single atom, which is not immersed in a \BEC.  While
this atom alone without a surrounding \BEC would not exhibit a phase
term in practice, it is possible to include this phase term in the
numerical calculation to illuminate its effect.  We will study the
full two-species system in the next paragraph.  We have seen in the
preceding sections that the critical phases $\alpha_\text{crit}$
correspond to a macroscopic jump in the momentum of the ground state.
As a consequence, if the induced phase crosses $\alpha_\text{crit}$,
the ground state of the impurity will jump to a higher momentum state.
Since the ToF distribution represents the momentum distribution of the
trapped atoms, we expect a central peak for nonrotating impurities and
a vanishing density at the center for impurities with a nonzero
momentum~\cite{Cozzini-PRA-2006,Nunnenkamp-PRA-2008}.  ToF images can
reveal this jump in the momentum of the impurity with good accuracy.
Corresponding density plots are shown in Fig.~\ref{fig:ToF}(a) for a
phase below and in Fig.~\ref{fig:ToF}(b) for a phase above the lowest
critical phase $\alpha_\text{crit} = 1/2N_s$.  In
Fig.~\ref{fig:ToF}(c) we plot the profiles of the density distribution
after a realistic time of flight.  The different profiles of the two
ToF images should be clearly distinguishable with typical camera
resolutions~\cite{NelLiWei07}.

\begin{figure}
  \centering%
  \includegraphics[width=\linewidth]{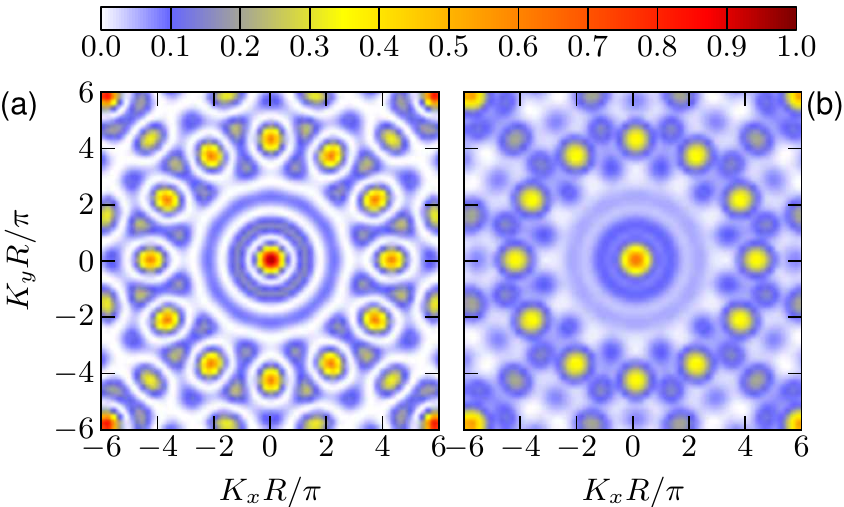}
  \caption{(Color online) Influence of temperature on the ToF images
    in the two-species Bose-Hubbard model. A single impurity is
    submerged into a \BEC with $6$ atoms in $N_s=12$ lattice sites,
    hopping $\tilde J_a=J_c$, interspecies coupling $U_I = 0.5 J_c$, \BEC
    interaction $U_c = 0.4 J_c$, and \BEC phase $\alpha_c = 0.042$.
    In (a) $k_BT = 0$ and in (b) $k_BT = 2J_c$.}\label{fig:ToF_mult}
\end{figure}

We now return to the numerical study of the full two-species
Bose-Hubbard model with a rotating \BEC and a stationary impurity
lattice.  For the moderately small system sizes studied here we cannot
expect to see a clear transition to a macroscopic occupation of a
nonzero impurity momentum state because the induced phase is not
sufficiently large to cause this jump.  Instead, in
Fig.~\ref{fig:ToF_mult} we show the influence of nonzero temperature
on the central peak in the ToF image of the impurity.  The \BEC phase
is chosen to be close to the critical phase $1/2N_s$ so that the
induced phase is expected to be large. The density distribution in
Fig.~\ref{fig:ToF_mult}(a) is very similar to the single-species
calculation in Fig.~\ref{fig:ToF}(a), which means that small
interspecies interactions do not influence the ToF image
significantly.  However, the induced phase is not sufficiently large
to cause a macroscopic occupation of a nonzero momentum state as
revealed by a vanishing central density in Fig.~\ref{fig:ToF}(b).
This is consistent with our analytical formula of the effective
polaron model, Eq.~\eqref{eq:alpha-lattice}, which does predict an induced
phase below the critical phase $1/2N_s$ for the parameters used in
Fig.~\ref{fig:ToF_mult}.  In Fig.~\ref{fig:ToF_mult}(b) we have chosen
the same parameters as in Fig.~\ref{fig:ToF_mult}(a) but with a large
temperature.  The finite temperature in this plot manifests itself in
a thermal background distribution superimposed on the coherent
distribution, which reduces the contrast of the ToF image.  However,
even at such large temperatures the features of the distribution are
still visible and we expect this method to be able to distinguish
between the distributions with a central peak or a central dip at
typical experimental temperatures.

An application of this phase detection method via ToF images might be
to use the impurity as a nondestructive probe to reveal the rotational
state of the underlying \BEC.  If the rotation is sufficiently close
to a critical phase, the probe will acquire a phase which is detected
by the central dip in the ToF image.

\section{Clustering of impurity atoms}\label{sec:clustering}
The presence of the coherent phonon cloud surrounding the lattice
atoms mediates a long-range interaction between distant atoms.
Furthermore, the onsite interaction decreases with the polaronic level
shift $E_p$.  For a sufficiently large mediated interaction this can
result in an attractive onsite potential~\cite{Klein-NJP-2007}.  In
this section, we will see how a phase twist influences the clustering
properties of the impurities.  All computations in this section are
based on numerically exact diagonalization of either the one- or
two-species Bose-Hubbard Hamiltonian.

\begin{figure}
  \centering
  \includegraphics[width=\linewidth]{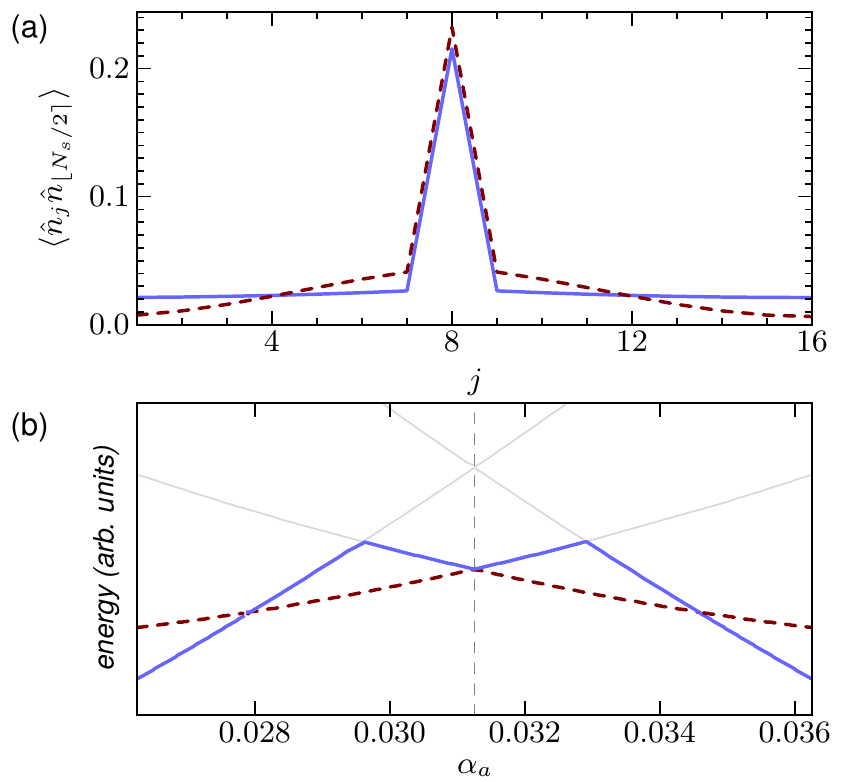}
  \caption{(Color online) (a) Correlations of impurities for $N_s =
    16$ lattice sites, $3$ atoms, and $U_a = -0.125\tilde J_a$. The
    two lines represent the two families of correlations depending on
    the momentum state of the system, i.e., the phase $\alpha_a$. Each
    family is a set of correlations for many values of $\alpha_a$.
    (b) The spectrum exhibits crossings below and above the critical
    phases $\alpha_\text{crit}$ (the first critical phase is indicated
    with a vertical dashed line), where the impurities switch their
    momentum ground state (thick solid and dashed lines). The thin
    gray lines indicate higher lying energy
    levels.}\label{fig:correlation}
\end{figure}

We study the effective single-species Bose-Hubbard model,
Eq.~\eqref{eq:H_eff}, which contains a phase term $\alpha_{j,j'} =
\alpha_a$ in the hopping, and assume an attractive interaction
($U_a/\tilde J_a < 0$).  The effect of the mediated interaction is
measured by the density-density correlation $\langle \hat n_j \hat
n_{j'}\rangle$ between sites $j$ and $j'$.  In
Fig.~\ref{fig:correlation}(a) we see the effect of the phase term on
the density-density correlations $\langle \hat n_j \hat n_{\lfloor
  N_s/2\rceil} \rangle$.  Depending on the value of the induced phase
the ground state is either in a weakly or a more strongly bound state.
The strongly bound state with a higher onsite correlation is observed
when the ground state of the system changes its momentum.  In the
energy spectrum this switch corresponds to a crossing of the two
lowest energy levels.  The energy spectrum of the system in
Fig.~\ref{fig:correlation}(b) shows such crossings close to the first
critical phase $\alpha_\text{crit} = 1/2N_s$.  We find two such
crossings below and above each critical phase.  For the values of
$\alpha_a$ in between two such crossings the ground state becomes
stronger bound.  Thus the two families of correlations indicated in
Fig.~\ref{fig:correlation}(a) consist of the correlations for all
$\alpha_a$ either between two crossing above and below a critical
phase or outside the crossing region.  For different sets of
parameters the picture may feature more than two differently bound
sets of states depending on the number of crossings within the lowest
energy level.  In conclusion, the effect of a phase term in the
attractive Bose-Hubbard model is to introduce energy crossings and to
bind the atoms more strongly for certain values of the phase.

\section{Quantum chaos}\label{sec:chaos}
We now briefly address the question if our system exhibits quantum
chaotic behavior as observed recently for a single-mode Bose-Hubbard
model with a phase twist~\cite{Buonsante-PRA-2008}.  In our
two-component Bose-Hubbard model, the phase of the second component is
induced through coupling with the first component.  This is different
from the one-species model, where such an interspecies coupling is not
present and where a phase term has to be assumed a priori.  To
formalize the notion of quantum chaos we denote with $\Delta E_j$ the
distance between two neighboring energy levels $E_j$ and $E_{j+1}$,
and with $\overline{\Delta E}$ the average over all spacings $\Delta
E_j$.  Furthermore, we define a quasi-continuous parameter $s = \Delta
E/\overline{\Delta E}$ for the normalized energy spacings.  Then a
nonchaotic system follows the Poisson distribution with a probability
density $p(s) = \exp(-s)$.  On the other hand, a quantum chaotic
system follows a \GOE distribution $p(s) = \pi s\exp(-\pi s^2/4)/2$.
The repulsion of levels expressed in the \GOE is a result of
correlations in the system, which are not present in the uncorrelated
Poissonian (nonchaotic) statistics.

\begin{figure}
  \centering
  \includegraphics[width=\linewidth]{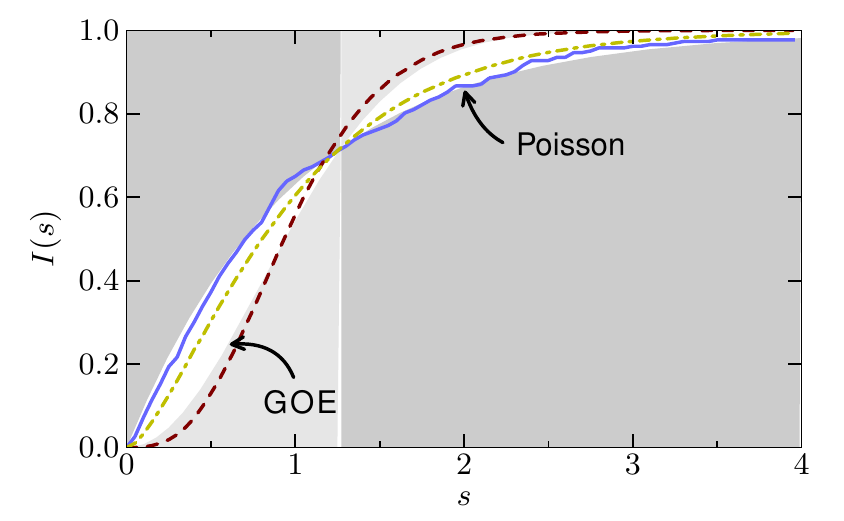}
  \caption{(Color online) Cumulative level spacing for different
    disorder strengths $e/J_c = 0, 1, 20$ (solid blue, dashed red,
    dashed-dotted yellow lines, respectively).  The edge of the dark
    shaded area indicates the Poisson prediction and the light shaded
    area the \GOE prediction.  The curves describe a crossover from
    regular behavior to chaotic, and back to regular for increasing
    disorder.  This is for $3$ \BEC atoms and $2$ impurities in
    lattices with $5$ sites averaged over $100$ independent disorder
    realizations.  The other parameters are $U_c = 1.5J_c$, $U_a =
    0.1J_c$, $U_I = J_c$, $\alpha_c = 0.057$.}\label{fig:chaos}
\end{figure}

\begin{figure}
  \centering
  \includegraphics[width=\linewidth]{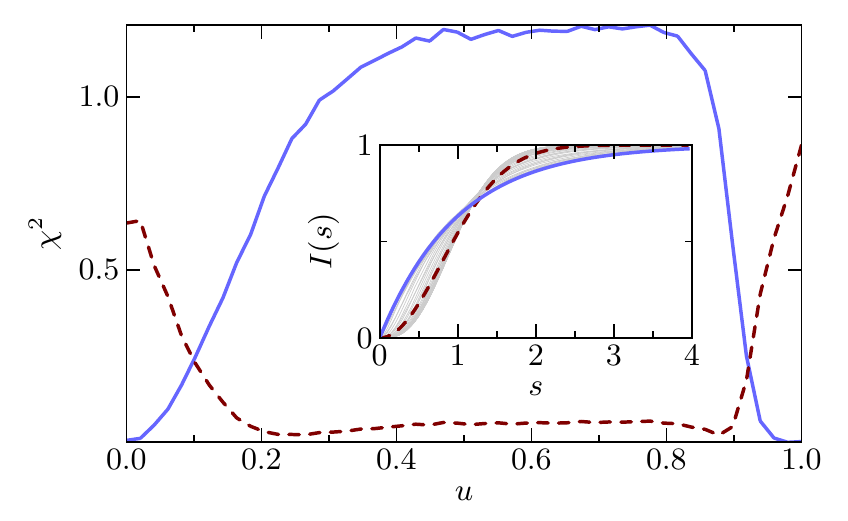}
  \caption{(Color online) $\chi^2$ tests of the Poisson (solid line)
    and \GOE predictions (dashed line).  For intermediate values of
    $u$ ($0.2 \alt u \alt 0.9$) the energy distribution closely
    follows the \GOE prediction.  Below and above these values we
    observe a more regular behavior as the system is in either the
    hopping-only or the no-hopping state, respectively.  As parameters
    we have chosen $N_s = 5$ lattice sites, $3$ \BEC atoms, $2$
    impurities, $\alpha_c = 0.057$.  The results have been averaged
    over $100$ independent disorder realizations.  The inset shows the
    individual cumulative level distributions (thin lines, which
    appear as a gray area at this resolution) together with the
    Poisson (solid line) and \GOE predictions (dashed
    line).}\label{fig:chaos2}
\end{figure}

Because of the various symmetries in the Bose-Hubbard model one cannot
expect a global quantum chaotic behavior of such a system.  One
approach is to restrict oneself to the local behavior in suitable
subspaces which do not exhibit these symmetries~\cite{KolBuc04}.
Another approach---and typically easier to achieve in experiments---is
to break symmetries explicitly, for example, breaking translational
symmetry by introducing random disorder~\cite{Buonsante-PRA-2008}.  We
introduce disorder into the two-component Bose-Hubbard Hamiltonian
with a phase twist by adding a local term with a random strength
\begin{equation*}
  \hat H_\text{dis} = \sum_j \left(e_j^{\vphantom{\dagger}} \hat c_j^\dagger \hat
    c_j^{\phantom{\dagger}} + e_j' \hat a_j^\dagger \hat
    a_j^{\phantom{\dagger}}\right).
\end{equation*}
The energies $e_j$ and $e_j'$ are two independent random variables
uniformly distributed on an interval $(-e/2, e/2)$.  In
Fig.~\ref{fig:chaos} we plot the integrated level distribution $I(s) =
\int_0^s p(s')\di s'$ for different values of the disorder strength
$e$.  For vanishing disorder, the system follows closely the regular
behavior of the Poisson distribution.  Choosing the same order of
magnitude for all energy scales $J_{a,c} \simeq U_{a,c} \simeq U_I
\simeq e$ we recover a similar behavior as observed in the
one-component Bose-Hubbard model in~\cite{Buonsante-PRA-2008}.  The
energy level distribution follows most closely the distribution of a
\GOE, which indicates quantum chaos.  Increasing the disorder further
reestablishes the regular behavior because the atoms tend to localize
in the disorder potential.  The effect of the interspecies interaction
on the level spacing is investigated in Fig.~\ref{fig:chaos2}.  There
we have defined a parameter $u = U_I/e = U_{a,c}/e$ with $J_{a,c}/e =
1-u$.  As we sweep $u$ from $0$ to $1$, the underlying Hamiltonian
changes from describing an ideal two-component system with hopping to
a fully interacting system without hopping.  In the absence of the
disorder potential these two extremal cases would lead to degenerate
eigenenergies (going from momentum Fock eigenstates to spatial Fock
states).  Disorder lifts the degeneracy and we can observe global
quantum chaotic behavior.  We recover very similar curves with the
definition $u = U_I/e$ and $J_{a,c}/e = U_{a,c}/e = 1-u$.  The
$\chi^2$ tests plotted in Fig.~\ref{fig:chaos2} indicate that the
system initially follows a regular behavior but quickly changes to a
\GOE distribution. Increasing the interaction further, ultimately the
behavior becomes regular again in the no-hopping regime.  We have
defined $\chi^2 = \sum_j [I_\text{num}(s_j) - I(s_j)]^2$, where
$I_\text{num}(s_j)$ are the numerically obtained cumulative level
spacings at point $s_j$ and $I(s_j)$ is either the integrated Poisson
or \GOE distribution.

\section{Conclusion}\label{sec:conclusion}
We have presented a method for creating an artificial magnetic field
in a ring of trapped neutral atoms.  Our method works by submerging
atoms trapped in an optical lattice into a \BEC which exhibits a phase
term.  We have discussed different setups which lead to an induced
phase in the impurities.  First, with a \BEC in a rotating ring,
second, with a rotating \BEC in a static ring.  We have then derived
an effective polaron model, which reduces the problem to a
single-species Hubbard model with a phase term on the hopping.  For
realistic parameters our analytical formulas predict induced phases up
to $\alpha_a \simeq 0.03$ in the first setup or $|\alpha_a| \simeq
0.25$ with a \BEC in an excited rotational state.  Comparisons with
numerical solutions of the full problem for small systems show that
our model can be used to qualitatively describe the system even beyond
the constraints set by the analytical derivation.  Furthermore, we
have discussed methods for observing the induced phase in the
impurities: by using ToF images and measuring mass currents.  The ToF
images show a sharp transition when a critical value of the induced
phase is crossed.  Increasing temperature or interspecies coupling
obscures the sharp transition but the main features of the transition
remain observable.  Finally, we have compared and extended studies of
quantum chaos to the two-species Bose-Hubbard Hamiltonian with a phase
twist.  By introducing random disorder we could observe the onset of
chaos when all energy scales in the system are of the same order,
similar to the result of a one-species Bose-Hubbard Hamiltonian with a
phase twist.  In the two-species model the interspecies coupling
ensures that the impurities exhibit a nontrivial phase, which is
necessary to observe quantum chaos in this model.

Contemporary optical lattice technology allows the investigation of
large systems of coherent neutral atoms evolving under a precisely
known Hamiltonian, which is often not the case in actual condensed
matter experiments.  The methods presented here can be seen as a
simulation of the effects of magnetic fields on ring systems.
However, the 1D treatment also allows us to gain insight into a 2D
system of impurities immersed into a rotating 2D \BEC~\cite{KleJak09}.
Our results suggest that, in general, one may expect a decrease in the
induced phase for nonvanishing interaction in 2D.  However, the
reduction is not expected to destroy the large phases obtainable in
2D.  Furthermore, detection methods similar to the ones presented here
will be applicable in the 2D setup.  In 2D a phase in the hopping term
of the Bose-Hubbard Hamiltonian gives rise to the quantum Hall
effect~\cite{Palmer-PRL-2006}.  In light of this prospect, an
extension of our methods to 2D will be of great interest and offer a
quantum simulator for even more condensed matter models.

\begin{acknowledgments}
  The authors thank S. R. Clark for fruitful discussions.  This
  research was supported by the European Commission under the Marie
  Curie Program through \caps{QIPEST}, by the United Kingdom EPSRC
  through QIP IRC (Grant No.  GR/S82176/01), EuroQUAM Project
  No. EP/E041612/1 and by the Keble Association (AK).
\end{acknowledgments}

\appendix*
\section{Bogoliubov approximation in the lattice with a phase twist}
To derive the Bogoliubov excitations for a \BEC in a 1D lattice with a
phase term in the hopping we follow \cite{Oosten-PRA-2001}, where the
authors derive the Bogoliubov approximation for a Bose-Hubbard model
without a phase twist.  First, we include the chemical potential
$\mu_c^L$ in the Bose-Hubbard Hamiltonian, Eq.~\eqref{eq:H_BH},
\begin{equation}\label{eq:H_BH-mu}
\begin{split}
  \hat H_{BH} &= -J_c \sum_j \left( \eu^{2\pi\im\alpha_c} \hat
    c_{j+1}^\dagger\hat c_j^{\phantom{\dagger}} +
    \eu^{-2\pi\im\alpha_c} \hat c_j^\dagger\hat c_{j+1}^{\phantom{\dagger}} \right)\\
  &\quad + \frac{U_c}{2} \sum_j^{\phantom{\dagger}} \hat c_j^\dagger
  \hat c_j^{\phantom{\dagger}}(\hat c_j^\dagger \hat
  c_j^{\phantom{\dagger}} - 1) - \mu_c^L \sum_j \hat c_j^\dagger \hat
  c_j^{\phantom{\dagger}}.
\end{split}
\end{equation}
We rewrite this Hamiltonian in the momentum representation with $\hat c_j =
(1/\sqrt{N_s}) \sum_q \hat d_q \eu^{\im 2\pi qj/N_s}$.  Noting that
$\sum_j \eu^{\im 2\pi j(q-q')/N_s} = N_s\delta_{q,q'}$ we arrive at
\begin{equation*}
\begin{split}
  \hat H_{BH} &= \sum_q \left(\bar\epsilon_q^L - \mu_c^L\right) \hat
  d_q^\dagger \hat d_q^{\phantom{\dagger}}\\
  &\quad + \frac{U_c}{2N_s} \sum_{q,q',q'',q'''} \hat d_q^\dagger \hat
  d_{q'}^\dagger \hat d_{q''}^{\phantom{\dagger}} \hat
  d_{q'''}^{\phantom{\dagger}} \delta_{q+q', q''+q'''},
\end{split}
\end{equation*}
where $\bar \epsilon_q^L = -2J_c\cos(2\pi q/N_s - \theta)$ and $\theta
= 2\pi\alpha_c$.  For the case considered in
Sec.~\ref{ssec:BEC-lattice} the \BEC is assumed to be in the ground
state with integer winding number $q_0$, which minimizes the energy
$\bar\epsilon_q^L$.  We therefore write $\theta = 2\pi q_0/N_s +
\Delta\theta/N_s$, where $\Delta\theta \in [-\pi, \pi)$ determines the
mismatch of the externally given phase twist $\theta$ and the phase of
the ground state.  The ground state energy is then given by
$\bar\epsilon_{q_0}^L = -2J_c \cos(\Delta\theta/N_s)$.

The ground state is occupied with a macroscopically large number of
atoms in mode $q_0$, that is $\langle \hat d_{q_0}^\dagger \hat
d_{q_0}^{\phantom{\dagger}} \rangle \simeq \langle \hat
d_{q_0}^{\phantom{\dagger}} \hat d_{q_0}^\dagger \rangle \sim N_0$ and
all other momentum state densities with $q \neq q_0$ are negligible.
This allows us to rewrite the creation and annihilation operators in
terms of a mean field contribution and quantized fluctuations, that is
\begin{align*}
  \hat d_{q_0}, \hat d_{q_0}^\dagger &\rightarrow \sqrt{N_0},\\
  \hat d_q^{\phantom{\dagger}}, \hat d_q^\dagger &\rightarrow \hat
  d_q^{\phantom{\dagger}}, \hat d_q^\dagger \quad \text{for } q \neq
  q_0.
\end{align*}
The chemical potential attains a value such that the Hamiltonian is
minimal with respect to $N_0$. It is given by $\mu_c^L =
\bar\epsilon_{q_0}^L + U_c\bar n_0$, where $\bar n_0 = N_0/N_s$ is the
density of the condensed atoms.  An effective Hamiltonian is derived
by only keeping terms up to second order in the fluctuations. For the
second order part we keep two operators and replace the respective
other two by $N_0$ in the interaction terms of the Hamiltonian.
Furthermore, we substitute the chemical potential and use the bosonic
commutation relation $[\hat d_q^{\phantom{\dagger}}, \hat d_q^\dagger]
= 1$.  This procedure yields the Hamiltonian
\begin{equation*}
\begin{split}
  \hat H^{(0+2)} &= -\frac{1}{2} U_c \bar n_0 N_0 - \frac{1}{2}
  \sideset{}{'}\sum_q \left(\bar\epsilon_q^L - \bar\epsilon_{q_0}^L +
    U_c\bar n_0\right)\\
  &\quad + \frac{1}{2} \sideset{}{'}\sum_q \left(\hat d_{q_0+q}^\dagger,
  \hat d_{q_0-q}^{\phantom{\dagger}}\right) D \begin{pmatrix}
    \hat d_{q_0+q}\\
    \hat d_{q_0-q}^\dagger
  \end{pmatrix},
\end{split}
\end{equation*}
where
\begin{equation*}
  D = \begin{pmatrix}
    \tilde\epsilon_{q_0+q}^L + U_c \bar n_0 & U_c \bar n_0\\
    U_c \bar n_0 & \tilde\epsilon_{q_0-q}^L + U_c \bar n_0
  \end{pmatrix}.
\end{equation*}
Here we have defined $\tilde\epsilon_{q_0\pm q}^L :=
\bar\epsilon_{q_0\pm q}^L - \bar\epsilon_{q_0}^L$. The quadratic terms
are diagonalized by the Bogoliubov transformation
\begin{equation}\label{eq:Bogoliubov-trafo}
  \begin{pmatrix}
    \hat b_q\\
    \hat b_{-q}^\dagger
  \end{pmatrix} = \begin{pmatrix}
    u_q & v_q\\
    v_q^* & u_q^*
  \end{pmatrix} \begin{pmatrix}
    \hat d_{q_0+q}\\
    \hat d_{q_0-q}^\dagger
  \end{pmatrix} =: B \begin{pmatrix}
    \hat d_{q_0+q}\\
    \hat d_{q_0-q}^\dagger
  \end{pmatrix}.
\end{equation}
In order for $\hat b_q$ and $\hat b_q^\dagger$ to obey the bosonic
commutation relation the coefficients have to fulfill $|u_q|^2 -
|v_q|^2 = 1$.  Plugging this expansion into the effective Hamiltonian
and requiring that the resulting matrix $(B^{-1})^\dagger DB^{-1}$ is
diagonal we arrive at the following conditions for the coefficients
\begin{align*}
  \tilde\epsilon_{q_0+q}^L |u_q|^2 +\tilde\epsilon_{q_0-q}^L|v_q|^2
  + U_c\bar n_0 |u_q - v_q|^2 &= \hbar\omega_q^L,\\
  - 2\epsilon_q^L u_q v_q + U_c\bar n_0 (u_q - v_q)^2 &=0,
\end{align*}
where $\epsilon_q^L = (\tilde\epsilon_{q_0+q}^L +
\tilde\epsilon_{q_0-q}^L)/2 = 4J_c\sin^2(\pi q/N_s)
\cos(\Delta\theta/N_s)$.  These equations have the solution
\begin{gather*}
  \hbar\omega_q^L = E_q^L + \Lambda_q,\\
  |v_q|^2 = |u_q|^2 - 1 = \frac{1}{2}\left(\frac{\epsilon_q^L +
      U_c\bar n_0}{E_q^L} - 1\right),
\end{gather*}
where we have defined $E_q^L = \sqrt{\epsilon_q^L (\epsilon_q^L +
  2U_c\bar n_0)}$ and $\Lambda_q := (\tilde\epsilon_{q_0+q}^L -
\tilde\epsilon_{q_0-q}^L)/2 = -2J_c \sin(2\pi q/N_s)
\sin(\Delta\theta/N_s)$.  Thus the Hamiltonian up to second order is
\begin{equation*}
\begin{split}
  \hat H^{(0+2)} &= -\frac{1}{2} U_c\bar n_0^2 N_s + \frac{1}{2}
  \sideset{}{'}\sum_q \left[\hbar\omega_q^L - \left(\bar\epsilon_q^L -
  \bar\epsilon_{q_0}^L + U_c\bar n_0\right)\right]\\
  &\quad + \frac{1}{2} \sideset{}{'}\sum_q \hbar\omega_q^L \hat
  b_q^\dagger \hat b_q^{\phantom{\dagger}}.\raisetag{1.5\baselineskip}
\end{split}
\end{equation*}
For vanishing phase twist $\Lambda_q = 0$ and the excitations have an
energy given by $E_q^L$, which has the form of the free Bogoliubov
energy $E_q^B$ but with the free dispersion replaced by the dispersion
relation of the first band in the lattice.  For small $q$ we see that
$E_q^L$ is linear in the quasimomentum, which means that the
quasiparticles behave like phonons as in the continuous case.  For
higher $q$ the spectrum resembles massive particles in a lattice.
Turning on a phase twist results in an asymmetry in the spectrum
because $\Lambda_q$ is an odd function of $q$.  Hence, the system of
quasiparticles will prefer the branch of quasimomenta with lower
energies, which will result in a drift of quasiparticles.

\begin{figure}
\centering%
\includegraphics[width=\linewidth]{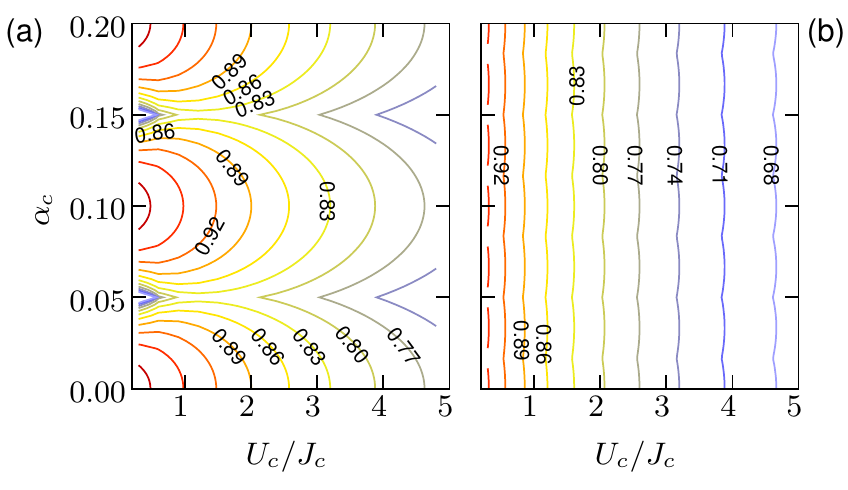}
\caption{(Color online) Contours of the relative condensate density
  $\bar n_0/\bar n$ at $T=0$ for (a) $N_s = 10$ and $\bar n = 0.5$,
  and (b) $N_s = 30$ and $\bar n = 2$.  The values in the plot
  indicate the relative density of the respective
  contour.}\label{fig:n0}
\end{figure}

Next we focus on the interaction Hamiltonian $\hat H_I$ between \BEC
and impurities, Eq.~\eqref{eq:H_I}.  As in the continuous case, we
only keep terms linear in the fluctuations in the interaction
Hamiltonian.  Rewriting it in the momentum representation and
introducing a macroscopic population of one mode $q_0$ yields
\begin{equation*}
\begin{split}
  H_I^{(0+1)} &= U_I\bar n_0 \sum_j \hat n_j\\
  &\quad + \frac{U_I\sqrt{N_0}}{N_s} \sum_{j} \hat n_j
  \sideset{}{'}\sum_q \Bigl(\hat d_{q+q_0}^\dagger \eu^{-\im 2\pi jq/N_s}\\
  &\qquad + \hat d_{q+q_0}^{\phantom{\dagger}} \eu^{\im 2\pi
    jq/N_s}\Bigr).
\end{split}
\end{equation*}
Here we have shifted the summation index by $q_0$ in order for the
Bogoliubov transformation, Eq.~\eqref{eq:Bogoliubov-trafo}, to be
applicable.  After application of the transformation we find
\begin{equation*}
\begin{split}
  H_I^{(0+1)} &= U_I\bar n_0 \sum_j \hat n_j\\
  &\quad + \sum_j \hat n_j \sideset{}{'}\sum_{q} \hbar\omega_q^L
  \left(M_{j,q}^L \hat b_q^{\phantom{\dagger}} + M_{j,q}^{L*} \hat
    b_q^\dagger \right),
\end{split}
\end{equation*}
where the coupling between the impurities and the Bogoliubov modes is
given by
\begin{equation*}
  M_{j,q}^L = \frac{U_I\sqrt{N_0}}{N_s}\frac{1}{\hbar\omega_q^L} (u_q^* -
  v_q^*)\, \eu^{-\im 2\pi jq/N_s}.
\end{equation*}

The condensate density $\bar n_0$ is defined as the macroscopic,
condensed part of the total density $\bar n = N_c/N_s$. The
noncondensed part then arises from the expectation value of the
fluctuation number operators, that is $\bar n = \bar n_0 +
\sideset{}{'}\sum_{q}\langle \hat d_q^\dagger \hat
d_q^{\phantom{\dagger}}\rangle/N_s$.  Substituting the fluctuations
via Eq.~\eqref{eq:Bogoliubov-trafo} we see that $\bar n_0$ is
implicitly given by
\begin{equation*}
\begin{split}
  \bar n &= \bar n_0 + \frac{1}{N_s} \sideset{}{'}\sum_q
  \Biggl[\frac{\epsilon_q^L + U_c \bar n_0}{2E_q^L} \Biggl(
  \frac{2}{\eu^{\hbar\omega_q^L/k_BT}-1} + 1 \Biggr) +
  \frac{1}{2} \Biggr].
\end{split}
\end{equation*}
The Boltzmann factor enters through averaging over the phonon
distribution at a temperature $T$.  In Fig.~\ref{fig:n0} we plot the
relative condensate density $\bar n_0/\bar n$ for two different
filling factors and lattice sizes.  It becomes clear that our
mean field approach is not valid close to the critical values of
$\alpha_c$ at very low interaction $U_c$ because the condensate
density vanishes [see Fig.~\ref{fig:n0}(a)].  On the other hand, we
see that for small interaction we can assume that $\bar n_0 \simeq
\bar n$ provided that $\alpha_c$ is not too close to a critical value.

\bibliography{simulationBEC}

\begin{thebibliography}{54}
\expandafter\ifx\csname natexlab\endcsname\relax\def\natexlab#1{#1}\fi
\expandafter\ifx\csname bibnamefont\endcsname\relax
  \def\bibnamefont#1{#1}\fi
\expandafter\ifx\csname bibfnamefont\endcsname\relax
  \def\bibfnamefont#1{#1}\fi
\expandafter\ifx\csname citenamefont\endcsname\relax
  \def\citenamefont#1{#1}\fi
\expandafter\ifx\csname url\endcsname\relax
  \def\url#1{\texttt{#1}}\fi
\expandafter\ifx\csname urlprefix\endcsname\relax\def\urlprefix{URL }\fi
\providecommand{\bibinfo}[2]{#2}
\providecommand{\eprint}[2][]{\url{#2}}

\bibitem[{\citenamefont{Lewenstein et~al.}(2007)\citenamefont{Lewenstein,
  Sanpera, Ahufinger, Damski, Sen, and Sen}}]{Lewenstein-AiP-2006}
\bibinfo{author}{\bibfnamefont{M.}~\bibnamefont{Lewenstein}},
  \bibinfo{author}{\bibfnamefont{A.}~\bibnamefont{Sanpera}},
  \bibinfo{author}{\bibfnamefont{V.}~\bibnamefont{Ahufinger}},
  \bibinfo{author}{\bibfnamefont{B.}~\bibnamefont{Damski}},
  \bibinfo{author}{\bibfnamefont{A.}~\bibnamefont{Sen}}, \bibnamefont{and}
  \bibinfo{author}{\bibfnamefont{U.}~\bibnamefont{Sen}}, \bibinfo{journal}{Adv.
  Phys.} \textbf{\bibinfo{volume}{56}}, \bibinfo{pages}{243}
  (\bibinfo{year}{2007}),
  \href{http://www.arxiv.org/abs/cond-mat/0606771}{\nolinkurl{arXiv:cond-mat/0%
606771}}.

\bibitem[{\citenamefont{Jaksch and Zoller}(2005)}]{Jaksch-Ann-2005}
\bibinfo{author}{\bibfnamefont{D.}~\bibnamefont{Jaksch}} \bibnamefont{and}
  \bibinfo{author}{\bibfnamefont{P.}~\bibnamefont{Zoller}},
  \bibinfo{journal}{Ann. Phys. (NY)} \textbf{\bibinfo{volume}{315}},
  \bibinfo{pages}{52} (\bibinfo{year}{2005}),
  \href{http://www.arxiv.org/abs/cond-mat/0410614}{\nolinkurl{arXiv:cond-mat/0%
410614}}.

\bibitem[{\citenamefont{{von Klitzing} et~al.}(1980)\citenamefont{{von
  Klitzing}, Dorda, and Pepper}}]{vonKlitzing-PRL-1980}
\bibinfo{author}{\bibfnamefont{K.}~\bibnamefont{{von Klitzing}}},
  \bibinfo{author}{\bibfnamefont{G.}~\bibnamefont{Dorda}}, \bibnamefont{and}
  \bibinfo{author}{\bibfnamefont{M.}~\bibnamefont{Pepper}},
  \bibinfo{journal}{Phys. Rev. Lett.} \textbf{\bibinfo{volume}{45}},
  \bibinfo{pages}{494} (\bibinfo{year}{1980}).

\bibitem[{\citenamefont{S{\o}rensen et~al.}(2005)\citenamefont{S{\o}rensen,
  Demler, and Lukin}}]{Sorensen-PRL-2005}
\bibinfo{author}{\bibfnamefont{A.~S.} \bibnamefont{S{\o}rensen}},
  \bibinfo{author}{\bibfnamefont{E.}~\bibnamefont{Demler}}, \bibnamefont{and}
  \bibinfo{author}{\bibfnamefont{M.~D.} \bibnamefont{Lukin}},
  \bibinfo{journal}{Phys. Rev. Lett.} \textbf{\bibinfo{volume}{94}},
  \bibinfo{pages}{086803} (\bibinfo{year}{2005}),
  \href{http://www.arxiv.org/abs/cond-mat/0405079}{\nolinkurl{arXiv:cond-mat/0%
405079}}.

\bibitem[{\citenamefont{Palmer and Jaksch}(2006)}]{Palmer-PRL-2006}
\bibinfo{author}{\bibfnamefont{R.~N.} \bibnamefont{Palmer}} \bibnamefont{and}
  \bibinfo{author}{\bibfnamefont{D.}~\bibnamefont{Jaksch}},
  \bibinfo{journal}{Phys. Rev. Lett.} \textbf{\bibinfo{volume}{96}},
  \bibinfo{pages}{180407} (\bibinfo{year}{2006}),
  \href{http://www.arxiv.org/abs/cond-mat/0604600v1}{\nolinkurl{arXiv:cond-mat%
/0604600v1}}.

\bibitem[{\citenamefont{Palmer et~al.}(2008)\citenamefont{Palmer, Klein, and
  Jaksch}}]{Palmer-PRA-2008}
\bibinfo{author}{\bibfnamefont{R.~N.} \bibnamefont{Palmer}},
  \bibinfo{author}{\bibfnamefont{A.}~\bibnamefont{Klein}}, \bibnamefont{and}
  \bibinfo{author}{\bibfnamefont{D.}~\bibnamefont{Jaksch}},
  \bibinfo{journal}{Phys. Rev. A} \textbf{\bibinfo{volume}{78}},
  \bibinfo{pages}{013609} (\bibinfo{year}{2008}),
  \href{http://www.arxiv.org/abs/0803.3771v2}{\nolinkurl{arXiv:0803.3771v2}}.

\bibitem[{\citenamefont{Schmidt}(1997)}]{SchMueUst97}
\bibinfo{author}{\bibfnamefont{V.~V.} \bibnamefont{Schmidt}},
  \emph{\bibinfo{title}{The physics of superconductors}}
  (\bibinfo{publisher}{Springer}, \bibinfo{address}{Berlin},
  \bibinfo{year}{1997}).

\bibitem[{\citenamefont{Ryu et~al.}(2007)\citenamefont{Ryu, Andersen, Clade,
  Natarajan, Helmerson, and Phillips}}]{RyuAndCla07}
\bibinfo{author}{\bibfnamefont{C.}~\bibnamefont{Ryu}},
  \bibinfo{author}{\bibfnamefont{M.~F.} \bibnamefont{Andersen}},
  \bibinfo{author}{\bibfnamefont{P.}~\bibnamefont{Clade}},
  \bibinfo{author}{\bibfnamefont{V.}~\bibnamefont{Natarajan}},
  \bibinfo{author}{\bibfnamefont{K.}~\bibnamefont{Helmerson}},
  \bibnamefont{and} \bibinfo{author}{\bibfnamefont{W.~D.}
  \bibnamefont{Phillips}}, \bibinfo{journal}{Phys.~Rev.~Lett.}
  \textbf{\bibinfo{volume}{99}}, \bibinfo{pages}{260401}
  (\bibinfo{year}{2007}),
  \href{http://www.arxiv.org/abs/0709.0012v2}{\nolinkurl{arXiv:0709.0012v2}}.

\bibitem[{\citenamefont{Wilkin and Gunn}(2000)}]{WilGun00}
\bibinfo{author}{\bibfnamefont{N.~K.} \bibnamefont{Wilkin}} \bibnamefont{and}
  \bibinfo{author}{\bibfnamefont{J.~M.~F.} \bibnamefont{Gunn}},
  \bibinfo{journal}{Phys.~Rev.~Lett.} \textbf{\bibinfo{volume}{84}},
  \bibinfo{pages}{6} (\bibinfo{year}{2000}),
  \href{http://www.arxiv.org/abs/cond-mat/9906282v1}{\nolinkurl{arXiv:cond-mat%
/9906282v1}}.

\bibitem[{\citenamefont{Cooper et~al.}(2001)\citenamefont{Cooper, Wilkin, and
  Gunn}}]{Cooper-PRL-2001}
\bibinfo{author}{\bibfnamefont{N.~R.} \bibnamefont{Cooper}},
  \bibinfo{author}{\bibfnamefont{N.~K.} \bibnamefont{Wilkin}},
  \bibnamefont{and} \bibinfo{author}{\bibfnamefont{J.~M.~F.}
  \bibnamefont{Gunn}}, \bibinfo{journal}{Phys. Rev. Lett.}
  \textbf{\bibinfo{volume}{87}}, \bibinfo{pages}{120405}
  (\bibinfo{year}{2001}),
  \href{http://www.arxiv.org/abs/cond-mat/0107005v1}{\nolinkurl{arXiv:cond-mat%
/0107005v1}}.

\bibitem[{\citenamefont{Schweikhard et~al.}(2004)\citenamefont{Schweikhard,
  Coddington, Engels, Mogendorff, and Cornell}}]{Schweikhard-PRL-2004}
\bibinfo{author}{\bibfnamefont{V.}~\bibnamefont{Schweikhard}},
  \bibinfo{author}{\bibfnamefont{I.}~\bibnamefont{Coddington}},
  \bibinfo{author}{\bibfnamefont{P.}~\bibnamefont{Engels}},
  \bibinfo{author}{\bibfnamefont{V.~P.} \bibnamefont{Mogendorff}},
  \bibnamefont{and} \bibinfo{author}{\bibfnamefont{E.~A.}
  \bibnamefont{Cornell}}, \bibinfo{journal}{Phys. Rev. Lett.}
  \textbf{\bibinfo{volume}{92}}, \bibinfo{pages}{040404}
  (\bibinfo{year}{2004}),
  \href{http://www.arxiv.org/abs/cond-mat/0308582v2}{\nolinkurl{arXiv:cond-mat%
/0308582v2}}.

\bibitem[{\citenamefont{Rosenbusch et~al.}(2002)\citenamefont{Rosenbusch,
  Petrov, Sinha, Chevy, Bretin, Castin, Shlyapnikov, and
  Dalibard}}]{RosPetSin02}
\bibinfo{author}{\bibfnamefont{P.}~\bibnamefont{Rosenbusch}},
  \bibinfo{author}{\bibfnamefont{D.~S.} \bibnamefont{Petrov}},
  \bibinfo{author}{\bibfnamefont{S.}~\bibnamefont{Sinha}},
  \bibinfo{author}{\bibfnamefont{F.}~\bibnamefont{Chevy}},
  \bibinfo{author}{\bibfnamefont{V.}~\bibnamefont{Bretin}},
  \bibinfo{author}{\bibfnamefont{Y.}~\bibnamefont{Castin}},
  \bibinfo{author}{\bibfnamefont{G.}~\bibnamefont{Shlyapnikov}},
  \bibnamefont{and} \bibinfo{author}{\bibfnamefont{J.}~\bibnamefont{Dalibard}},
  \bibinfo{journal}{Phys.~Rev.~Lett.} \textbf{\bibinfo{volume}{88}},
  \bibinfo{pages}{250403} (\bibinfo{year}{2002}),
  \href{http://www.arxiv.org/abs/cond-mat/0201568v2}{\nolinkurl{arXiv:cond-mat%
/0201568v2}}.

\bibitem[{\citenamefont{Bretin et~al.}(2004)\citenamefont{Bretin, Stock,
  Seurin, and Dalibard}}]{Bretin-PRL-2004}
\bibinfo{author}{\bibfnamefont{V.}~\bibnamefont{Bretin}},
  \bibinfo{author}{\bibfnamefont{S.}~\bibnamefont{Stock}},
  \bibinfo{author}{\bibfnamefont{Y.}~\bibnamefont{Seurin}}, \bibnamefont{and}
  \bibinfo{author}{\bibfnamefont{J.}~\bibnamefont{Dalibard}},
  \bibinfo{journal}{Phys. Rev. Lett.} \textbf{\bibinfo{volume}{92}},
  \bibinfo{pages}{050403} (\bibinfo{year}{2004}),
  \href{http://www.arxiv.org/abs/cond-mat/0307464v1}{\nolinkurl{arXiv:cond-mat%
/0307464v1}}.

\bibitem[{\citenamefont{Juzeli\={u}nas and
  {\"O}hberg}(2004)}]{Juzeliunas-PRL-2004}
\bibinfo{author}{\bibfnamefont{G.}~\bibnamefont{Juzeli\={u}nas}}
  \bibnamefont{and}
  \bibinfo{author}{\bibfnamefont{P.}~\bibnamefont{{\"O}hberg}},
  \bibinfo{journal}{Phys. Rev. Lett.} \textbf{\bibinfo{volume}{93}},
  \bibinfo{pages}{033602} (\bibinfo{year}{2004}),
  \href{http://www.arxiv.org/abs/cond-mat/0402317v2}{\nolinkurl{arXiv:cond-mat%
/0402317v2}}.

\bibitem[{\citenamefont{Juzeli\={u}nas
  et~al.}(2005)\citenamefont{Juzeli\={u}nas, {\"O}hberg, Ruseckas, and
  Klein}}]{Juzeliunas-PRA-2005}
\bibinfo{author}{\bibfnamefont{G.}~\bibnamefont{Juzeli\={u}nas}},
  \bibinfo{author}{\bibfnamefont{P.}~\bibnamefont{{\"O}hberg}},
  \bibinfo{author}{\bibfnamefont{J.}~\bibnamefont{Ruseckas}}, \bibnamefont{and}
  \bibinfo{author}{\bibfnamefont{A.}~\bibnamefont{Klein}},
  \bibinfo{journal}{Phys. Rev. A} \textbf{\bibinfo{volume}{71}},
  \bibinfo{pages}{053614} (\bibinfo{year}{2005}),
  \href{http://www.arxiv.org/abs/cond-mat/0412015v3}{\nolinkurl{arXiv:cond-mat%
/0412015v3}}.

\bibitem[{\citenamefont{Jaksch and Zoller}(2003)}]{Jaksch-NJP-2003}
\bibinfo{author}{\bibfnamefont{D.}~\bibnamefont{Jaksch}} \bibnamefont{and}
  \bibinfo{author}{\bibfnamefont{P.}~\bibnamefont{Zoller}},
  \bibinfo{journal}{New J. Phys.} \textbf{\bibinfo{volume}{5}},
  \bibinfo{pages}{56} (\bibinfo{year}{2003}),
  \href{http://www.arxiv.org/abs/quant-ph/0304038v1}{\nolinkurl{arXiv:quant-ph%
/0304038v1}}.

\bibitem[{\citenamefont{Klein and Jaksch}(2009)}]{KleJak09}
\bibinfo{author}{\bibfnamefont{A.}~\bibnamefont{Klein}} \bibnamefont{and}
  \bibinfo{author}{\bibfnamefont{D.}~\bibnamefont{Jaksch}},
  \bibinfo{journal}{Europhys.~Lett.} \textbf{\bibinfo{volume}{85}},
  \bibinfo{pages}{13001} (\bibinfo{year}{2009}),
  \href{http://www.arxiv.org/abs/0808.1898}{\nolinkurl{arXiv:0808.1898}}.

\bibitem[{\citenamefont{Morizot et~al.}(2006)\citenamefont{Morizot, Colombe,
  Lorent, Perrin, and Garraway}}]{MorColLor06}
\bibinfo{author}{\bibfnamefont{O.}~\bibnamefont{Morizot}},
  \bibinfo{author}{\bibfnamefont{Y.}~\bibnamefont{Colombe}},
  \bibinfo{author}{\bibfnamefont{V.}~\bibnamefont{Lorent}},
  \bibinfo{author}{\bibfnamefont{H.}~\bibnamefont{Perrin}}, \bibnamefont{and}
  \bibinfo{author}{\bibfnamefont{B.~M.} \bibnamefont{Garraway}},
  \bibinfo{journal}{Phys.~Rev.~A} \textbf{\bibinfo{volume}{74}},
  \bibinfo{pages}{023617} (\bibinfo{year}{2006}).

\bibitem[{\citenamefont{Gupta et~al.}(2005)\citenamefont{Gupta, Murch, Moore,
  Purdy, and Stamper-Kurn}}]{Gupta-PRL-2005}
\bibinfo{author}{\bibfnamefont{S.}~\bibnamefont{Gupta}},
  \bibinfo{author}{\bibfnamefont{K.~W.} \bibnamefont{Murch}},
  \bibinfo{author}{\bibfnamefont{K.~L.} \bibnamefont{Moore}},
  \bibinfo{author}{\bibfnamefont{T.~P.} \bibnamefont{Purdy}}, \bibnamefont{and}
  \bibinfo{author}{\bibfnamefont{D.~M.} \bibnamefont{Stamper-Kurn}},
  \bibinfo{journal}{Phys. Rev. Lett.} \textbf{\bibinfo{volume}{95}},
  \bibinfo{pages}{143201} (\bibinfo{year}{2005}),
  \href{http://www.arxiv.org/abs/cond-mat/0504749v1}{\nolinkurl{arXiv:cond-mat%
/0504749v1}}.

\bibitem[{\citenamefont{Heathcote et~al.}(2008)\citenamefont{Heathcote, Nugent,
  Sheard, and Foot}}]{Heathcote-NJP-2008}
\bibinfo{author}{\bibfnamefont{W.~H.} \bibnamefont{Heathcote}},
  \bibinfo{author}{\bibfnamefont{E.}~\bibnamefont{Nugent}},
  \bibinfo{author}{\bibfnamefont{B.~T.} \bibnamefont{Sheard}},
  \bibnamefont{and} \bibinfo{author}{\bibfnamefont{C.~J.} \bibnamefont{Foot}},
  \bibinfo{journal}{New J. Phys.} \textbf{\bibinfo{volume}{10}},
  \bibinfo{pages}{043012} (\bibinfo{year}{2008}).

\bibitem[{\citenamefont{Nandi et~al.}(2004)\citenamefont{Nandi, Walser, and
  Schleich}}]{NanWalSch04}
\bibinfo{author}{\bibfnamefont{G.}~\bibnamefont{Nandi}},
  \bibinfo{author}{\bibfnamefont{R.}~\bibnamefont{Walser}}, \bibnamefont{and}
  \bibinfo{author}{\bibfnamefont{W.~P.} \bibnamefont{Schleich}},
  \bibinfo{journal}{Phys.~Rev.~A} \textbf{\bibinfo{volume}{69}},
  \bibinfo{pages}{063606} (\bibinfo{year}{2004}).

\bibitem[{\citenamefont{Bruderer et~al.}(2007)\citenamefont{Bruderer, Klein,
  Clark, and Jaksch}}]{Bruderer-PRA-2007}
\bibinfo{author}{\bibfnamefont{M.}~\bibnamefont{Bruderer}},
  \bibinfo{author}{\bibfnamefont{A.}~\bibnamefont{Klein}},
  \bibinfo{author}{\bibfnamefont{S.~R.} \bibnamefont{Clark}}, \bibnamefont{and}
  \bibinfo{author}{\bibfnamefont{D.}~\bibnamefont{Jaksch}},
  \bibinfo{journal}{Phys. Rev. A} \textbf{\bibinfo{volume}{76}},
  \bibinfo{pages}{011605(R)} (\bibinfo{year}{2007}),
  \href{http://www.arxiv.org/abs/0704.2757v2}{\nolinkurl{arXiv:0704.2757v2}}.

\bibitem[{\citenamefont{Bruderer et~al.}(2008)\citenamefont{Bruderer, Klein,
  Clark, and Jaksch}}]{Bruderer-NJP-2008}
\bibinfo{author}{\bibfnamefont{M.}~\bibnamefont{Bruderer}},
  \bibinfo{author}{\bibfnamefont{A.}~\bibnamefont{Klein}},
  \bibinfo{author}{\bibfnamefont{S.~R.} \bibnamefont{Clark}}, \bibnamefont{and}
  \bibinfo{author}{\bibfnamefont{D.}~\bibnamefont{Jaksch}},
  \bibinfo{journal}{New J. Phys.} \textbf{\bibinfo{volume}{10}},
  \bibinfo{pages}{033015} (\bibinfo{year}{2008}),
  \href{http://www.arxiv.org/abs/0710.4493v2}{\nolinkurl{arXiv:0710.4493v2}}.

\bibitem[{\citenamefont{Klein et~al.}(2007)\citenamefont{Klein, Bruderer,
  Clark, and Jaksch}}]{Klein-NJP-2007}
\bibinfo{author}{\bibfnamefont{A.}~\bibnamefont{Klein}},
  \bibinfo{author}{\bibfnamefont{M.}~\bibnamefont{Bruderer}},
  \bibinfo{author}{\bibfnamefont{S.~R.} \bibnamefont{Clark}}, \bibnamefont{and}
  \bibinfo{author}{\bibfnamefont{D.}~\bibnamefont{Jaksch}},
  \bibinfo{journal}{New J. Phys.} \textbf{\bibinfo{volume}{9}},
  \bibinfo{pages}{411} (\bibinfo{year}{2007}),
  \href{http://www.arxiv.org/abs/0710.3539v2}{\nolinkurl{arXiv:0710.3539v2}}.

\bibitem[{\citenamefont{Nunnenkamp et~al.}(2008)\citenamefont{Nunnenkamp, Rey,
  and Burnett}}]{Nunnenkamp-PRA-2008}
\bibinfo{author}{\bibfnamefont{A.}~\bibnamefont{Nunnenkamp}},
  \bibinfo{author}{\bibfnamefont{A.~M.} \bibnamefont{Rey}}, \bibnamefont{and}
  \bibinfo{author}{\bibfnamefont{K.}~\bibnamefont{Burnett}},
  \bibinfo{journal}{Phys. Rev. A} \textbf{\bibinfo{volume}{77}},
  \bibinfo{pages}{023622} (\bibinfo{year}{2008}),
  \href{http://www.arxiv.org/abs/0711.3831v2}{\nolinkurl{arXiv:0711.3831v2}}.

\bibitem[{\citenamefont{Bohigas et~al.}(1984)\citenamefont{Bohigas, Giannoni,
  and Schmit}}]{BohGiaSch84}
\bibinfo{author}{\bibfnamefont{O.}~\bibnamefont{Bohigas}},
  \bibinfo{author}{\bibfnamefont{M.~J.} \bibnamefont{Giannoni}},
  \bibnamefont{and} \bibinfo{author}{\bibfnamefont{C.}~\bibnamefont{Schmit}},
  \bibinfo{journal}{Phys.~Rev.~Lett.} \textbf{\bibinfo{volume}{52}},
  \bibinfo{pages}{1} (\bibinfo{year}{1984}).

\bibitem[{\citenamefont{Kolovsky and Buchleitner}(2004)}]{KolBuc04}
\bibinfo{author}{\bibfnamefont{A.~R.} \bibnamefont{Kolovsky}} \bibnamefont{and}
  \bibinfo{author}{\bibfnamefont{A.}~\bibnamefont{Buchleitner}},
  \bibinfo{journal}{Europhys.~Lett.} \textbf{\bibinfo{volume}{68}},
  \bibinfo{pages}{632} (\bibinfo{year}{2004}),
  \href{http://www.arxiv.org/abs/cond-mat/0403213v2}{\nolinkurl{arXiv:cond-mat%
/0403213v2}}.

\bibitem[{\citenamefont{Mahan}(2000)}]{Mahan-2000}
\bibinfo{author}{\bibfnamefont{G.~D.} \bibnamefont{Mahan}},
  \emph{\bibinfo{title}{Many-{P}article {P}hysics}} (\bibinfo{publisher}{Kluwer
  {A}cademic/Plenum Publishers}, \bibinfo{address}{New {Y}ork},
  \bibinfo{year}{2000}).

\bibitem[{\citenamefont{Buonsante and Wimberger}(2008)}]{Buonsante-PRA-2008}
\bibinfo{author}{\bibfnamefont{P.}~\bibnamefont{Buonsante}} \bibnamefont{and}
  \bibinfo{author}{\bibfnamefont{S.}~\bibnamefont{Wimberger}},
  \bibinfo{journal}{Phys. Rev. A} \textbf{\bibinfo{volume}{77}},
  \bibinfo{pages}{041606(R)} (\bibinfo{year}{2008}),
  \href{http://www.arxiv.org/abs/0710.1853v3}{\nolinkurl{arXiv:0710.1853v3}}.

\bibitem[{\citenamefont{{Franke-Arnold}
  et~al.}(2007)\citenamefont{{Franke-Arnold}, Leach, Padgett, Lembessis,
  Ellinas, Wright, Girkin, {\" O}hberg, and Arnold}}]{FraLeaPad07}
\bibinfo{author}{\bibfnamefont{S.}~\bibnamefont{{Franke-Arnold}}},
  \bibinfo{author}{\bibfnamefont{J.}~\bibnamefont{Leach}},
  \bibinfo{author}{\bibfnamefont{M.~J.} \bibnamefont{Padgett}},
  \bibinfo{author}{\bibfnamefont{V.~E.} \bibnamefont{Lembessis}},
  \bibinfo{author}{\bibfnamefont{D.}~\bibnamefont{Ellinas}},
  \bibinfo{author}{\bibfnamefont{A.~J.} \bibnamefont{Wright}},
  \bibinfo{author}{\bibfnamefont{J.~M.} \bibnamefont{Girkin}},
  \bibinfo{author}{\bibfnamefont{P.}~\bibnamefont{{\" O}hberg}},
  \bibnamefont{and} \bibinfo{author}{\bibfnamefont{A.~S.}
  \bibnamefont{Arnold}}, \bibinfo{journal}{Opt. Express}
  \textbf{\bibinfo{volume}{15}}, \bibinfo{pages}{8619} (\bibinfo{year}{2007}).

\bibitem[{\citenamefont{LeBlanc and Thywissen}(2007)}]{Leblanc-PRA-2007}
\bibinfo{author}{\bibfnamefont{L.~J.} \bibnamefont{LeBlanc}} \bibnamefont{and}
  \bibinfo{author}{\bibfnamefont{J.~H.} \bibnamefont{Thywissen}},
  \bibinfo{journal}{Phys. Rev. A} \textbf{\bibinfo{volume}{75}},
  \bibinfo{pages}{053612} (\bibinfo{year}{2007}).

\bibitem[{\citenamefont{Petrov et~al.}(2000)\citenamefont{Petrov, Shlyapnikov,
  and Walraven}}]{PetShlWal00}
\bibinfo{author}{\bibfnamefont{D.~S.} \bibnamefont{Petrov}},
  \bibinfo{author}{\bibfnamefont{G.~V.} \bibnamefont{Shlyapnikov}},
  \bibnamefont{and} \bibinfo{author}{\bibfnamefont{J.~T.~M.}
  \bibnamefont{Walraven}}, \bibinfo{journal}{Phys.~Rev.~Lett.}
  \textbf{\bibinfo{volume}{85}}, \bibinfo{pages}{3745} (\bibinfo{year}{2000}),
  \href{http://www.arxiv.org/abs/cond-mat/0006339v1}{\nolinkurl{arXiv:cond-mat%
/0006339v1}}.

\bibitem[{\citenamefont{Castin}(2004)}]{Cas04}
\bibinfo{author}{\bibfnamefont{Y.}~\bibnamefont{Castin}}, \bibinfo{journal}{J.
  Phys. {IV} France} \textbf{\bibinfo{volume}{116}}, \bibinfo{pages}{89}
  (\bibinfo{year}{2004}),
  \href{http://www.arxiv.org/abs/cond-mat/0407118v2}{\nolinkurl{arXiv:cond-mat%
/0407118v2}}.

\bibitem[{\citenamefont{Baym et~al.}(1999)\citenamefont{Baym, Blaizot,
  Holzmann, Lalo{\" e}, and Vautherin}}]{BayBlaHol99}
\bibinfo{author}{\bibfnamefont{G.}~\bibnamefont{Baym}},
  \bibinfo{author}{\bibfnamefont{J.-P.} \bibnamefont{Blaizot}},
  \bibinfo{author}{\bibfnamefont{M.}~\bibnamefont{Holzmann}},
  \bibinfo{author}{\bibfnamefont{F.}~\bibnamefont{Lalo{\" e}}},
  \bibnamefont{and}
  \bibinfo{author}{\bibfnamefont{D.}~\bibnamefont{Vautherin}},
  \bibinfo{journal}{Phys.~Rev.~Lett.} \textbf{\bibinfo{volume}{83}},
  \bibinfo{pages}{1703} (\bibinfo{year}{1999}),
  \href{http://www.arxiv.org/abs/cond-mat/9905430v3}{\nolinkurl{arXiv:cond-mat%
/9905430v3}}.

\bibitem[{\citenamefont{Zobay and Rosenkranz}(2006)}]{ZobRos06}
\bibinfo{author}{\bibfnamefont{O.}~\bibnamefont{Zobay}} \bibnamefont{and}
  \bibinfo{author}{\bibfnamefont{M.}~\bibnamefont{Rosenkranz}},
  \bibinfo{journal}{Phys.~Rev.~A} \textbf{\bibinfo{volume}{74}},
  \bibinfo{pages}{053623} (\bibinfo{year}{2006}).

\bibitem[{\citenamefont{Kohn}(1959)}]{Koh59}
\bibinfo{author}{\bibfnamefont{W.}~\bibnamefont{Kohn}}, \bibinfo{journal}{Phys.
  Rev.} \textbf{\bibinfo{volume}{115}}, \bibinfo{pages}{809}
  (\bibinfo{year}{1959}).

\bibitem[{\citenamefont{Jaksch et~al.}(1998)\citenamefont{Jaksch, Bruder,
  Cirac, Gardiner, and Zoller}}]{Jaksch-PRL-1998}
\bibinfo{author}{\bibfnamefont{D.}~\bibnamefont{Jaksch}},
  \bibinfo{author}{\bibfnamefont{C.}~\bibnamefont{Bruder}},
  \bibinfo{author}{\bibfnamefont{J.~I.} \bibnamefont{Cirac}},
  \bibinfo{author}{\bibfnamefont{C.~W.} \bibnamefont{Gardiner}},
  \bibnamefont{and} \bibinfo{author}{\bibfnamefont{P.}~\bibnamefont{Zoller}},
  \bibinfo{journal}{Phys. Rev. Lett.} \textbf{\bibinfo{volume}{81}},
  \bibinfo{pages}{3108} (\bibinfo{year}{1998}),
  \href{http://www.arxiv.org/abs/cond-mat/9805329v3}{\nolinkurl{arXiv:cond-mat%
/9805329v3}}.

\bibitem[{\citenamefont{Holstein}(1959)}]{Holstein-Ann-1959}
\bibinfo{author}{\bibfnamefont{T.}~\bibnamefont{Holstein}},
  \bibinfo{journal}{Annals of Physics (NY)} \textbf{\bibinfo{volume}{8}},
  \bibinfo{pages}{343} (\bibinfo{year}{1959}).

\bibitem[{\citenamefont{Madison et~al.}(2001)\citenamefont{Madison, Chevy,
  Bretin, and Dalibard}}]{MadCheBre01}
\bibinfo{author}{\bibfnamefont{K.~W.} \bibnamefont{Madison}},
  \bibinfo{author}{\bibfnamefont{F.}~\bibnamefont{Chevy}},
  \bibinfo{author}{\bibfnamefont{V.}~\bibnamefont{Bretin}}, \bibnamefont{and}
  \bibinfo{author}{\bibfnamefont{J.}~\bibnamefont{Dalibard}},
  \bibinfo{journal}{Phys.~Rev.~Lett.} \textbf{\bibinfo{volume}{86}},
  \bibinfo{pages}{4443} (\bibinfo{year}{2001}),
  \href{http://www.arxiv.org/abs/cond-mat/0101051v1}{\nolinkurl{arXiv:cond-mat%
/0101051v1}}.

\bibitem[{\citenamefont{Hodby et~al.}(2001)\citenamefont{Hodby, Hechenblaikner,
  Hopkins, Marago, and Foot}}]{HodHecHop01}
\bibinfo{author}{\bibfnamefont{E.}~\bibnamefont{Hodby}},
  \bibinfo{author}{\bibfnamefont{G.}~\bibnamefont{Hechenblaikner}},
  \bibinfo{author}{\bibfnamefont{S.~A.} \bibnamefont{Hopkins}},
  \bibinfo{author}{\bibfnamefont{O.~M.} \bibnamefont{Marago}},
  \bibnamefont{and} \bibinfo{author}{\bibfnamefont{C.~J.} \bibnamefont{Foot}},
  \bibinfo{journal}{Phys.~Rev.~Lett.} \textbf{\bibinfo{volume}{88}},
  \bibinfo{pages}{010405} (\bibinfo{year}{2001}),
  \href{http://www.arxiv.org/abs/cond-mat/0106262v1}{\nolinkurl{arXiv:cond-mat%
/0106262v1}}.

\bibitem[{\citenamefont{Carusotto and Castin}(2004)}]{CarCas04}
\bibinfo{author}{\bibfnamefont{I.}~\bibnamefont{Carusotto}} \bibnamefont{and}
  \bibinfo{author}{\bibfnamefont{Y.}~\bibnamefont{Castin}},
  \bibinfo{journal}{C. R. Physique} \textbf{\bibinfo{volume}{5}},
  \bibinfo{pages}{107} (\bibinfo{year}{2004}),
  \href{http://www.arxiv.org/abs/cond-mat/0311601v2}{\nolinkurl{arXiv:cond-mat%
/0311601v2}}.

\bibitem[{\citenamefont{Rey et~al.}(2007)\citenamefont{Rey, Burnett, Satija,
  and Clark}}]{Rey-PRA-2007}
\bibinfo{author}{\bibfnamefont{A.~M.} \bibnamefont{Rey}},
  \bibinfo{author}{\bibfnamefont{K.}~\bibnamefont{Burnett}},
  \bibinfo{author}{\bibfnamefont{I.~I.} \bibnamefont{Satija}},
  \bibnamefont{and} \bibinfo{author}{\bibfnamefont{C.~W.} \bibnamefont{Clark}},
  \bibinfo{journal}{Phys. Rev. A} \textbf{\bibinfo{volume}{75}},
  \bibinfo{pages}{063616} (\bibinfo{year}{2007}),
  \href{http://www.arxiv.org/abs/cond-mat/0611332v1}{\nolinkurl{arXiv:cond-mat%
/0611332v1}}.

\bibitem[{\citenamefont{Fetter}(1972)}]{Fetter-AP-1972}
\bibinfo{author}{\bibfnamefont{A.~L.} \bibnamefont{Fetter}},
  \bibinfo{journal}{Ann.~Phys.} \textbf{\bibinfo{volume}{70}},
  \bibinfo{pages}{67} (\bibinfo{year}{1972}).

\bibitem[{\citenamefont{{\" O}gren and Kavoulakis}(2009)}]{OegKav09}
\bibinfo{author}{\bibfnamefont{M.}~\bibnamefont{{\" O}gren}} \bibnamefont{and}
  \bibinfo{author}{\bibfnamefont{G.}~\bibnamefont{Kavoulakis}},
  \bibinfo{journal}{J. Low Temp. Phys.} \textbf{\bibinfo{volume}{154}},
  \bibinfo{pages}{30} (\bibinfo{year}{2009}).

\bibitem[{\citenamefont{Landau}(1941)}]{Lan41}
\bibinfo{author}{\bibfnamefont{L.}~\bibnamefont{Landau}}, \bibinfo{journal}{J.
  Phys. U.S.S.R.} \textbf{\bibinfo{volume}{5}}, \bibinfo{pages}{71}
  (\bibinfo{year}{1941}).

\bibitem[{\citenamefont{Leggett}(1999)}]{Leg99}
\bibinfo{author}{\bibfnamefont{A.~J.} \bibnamefont{Leggett}},
  \bibinfo{journal}{Rev.~Mod.~Phys.} \textbf{\bibinfo{volume}{71}},
  \bibinfo{pages}{S318} (\bibinfo{year}{1999}).

\bibitem[{\citenamefont{Sykes et~al.}(2009)\citenamefont{Sykes, Davis, and
  Roberts}}]{SykDavRob09}
\bibinfo{author}{\bibfnamefont{A.~G.} \bibnamefont{Sykes}},
  \bibinfo{author}{\bibfnamefont{M.~J.} \bibnamefont{Davis}}, \bibnamefont{and}
  \bibinfo{author}{\bibfnamefont{D.~C.} \bibnamefont{Roberts}},
  \bibinfo{journal}{Phys.~Rev.~Lett.} \textbf{\bibinfo{volume}{103}},
  \bibinfo{pages}{085302} (\bibinfo{year}{2009}),
  \href{http://www.arxiv.org/abs/0904.0995}{\nolinkurl{arXiv:0904.0995}}.

\bibitem[{\citenamefont{Roberts and Pomeau}(2005)}]{RobPom05}
\bibinfo{author}{\bibfnamefont{D.~C.} \bibnamefont{Roberts}} \bibnamefont{and}
  \bibinfo{author}{\bibfnamefont{Y.}~\bibnamefont{Pomeau}},
  \bibinfo{journal}{Phys.~Rev.~Lett.} \textbf{\bibinfo{volume}{95}},
  \bibinfo{pages}{145303} (\bibinfo{year}{2005}).

\bibitem[{\citenamefont{Roberts}(2006)}]{Rob06}
\bibinfo{author}{\bibfnamefont{D.~C.} \bibnamefont{Roberts}},
  \bibinfo{journal}{Phys.~Rev.~A} \textbf{\bibinfo{volume}{74}},
  \bibinfo{pages}{013613} (\bibinfo{year}{2006}).

\bibitem[{\citenamefont{Kuper and Whitfield}(1963)}]{KupWhi63}
\bibinfo{editor}{\bibfnamefont{C.~G.} \bibnamefont{Kuper}} \bibnamefont{and}
  \bibinfo{editor}{\bibfnamefont{G.~D.} \bibnamefont{Whitfield}}, eds.,
  \emph{\bibinfo{title}{Polarons and Excitons}} (\bibinfo{publisher}{Oliver and
  Boyd}, \bibinfo{address}{Edinburgh}, \bibinfo{year}{1963}).

\bibitem[{\citenamefont{Bloch et~al.}(2008)\citenamefont{Bloch, Dalibard, and
  Zwerger}}]{Bloch-RMP-2008}
\bibinfo{author}{\bibfnamefont{I.}~\bibnamefont{Bloch}},
  \bibinfo{author}{\bibfnamefont{J.}~\bibnamefont{Dalibard}}, \bibnamefont{and}
  \bibinfo{author}{\bibfnamefont{W.}~\bibnamefont{Zwerger}},
  \bibinfo{journal}{Rev.~Mod.~Phys.} \textbf{\bibinfo{volume}{80}},
  \bibinfo{pages}{885} (\bibinfo{year}{2008}),
  \href{http://www.arxiv.org/abs/0704.3011v2}{\nolinkurl{arXiv:0704.3011v2}}.

\bibitem[{\citenamefont{Cozzini et~al.}(2006)\citenamefont{Cozzini, Jackson,
  and Stringari}}]{Cozzini-PRA-2006}
\bibinfo{author}{\bibfnamefont{M.}~\bibnamefont{Cozzini}},
  \bibinfo{author}{\bibfnamefont{B.}~\bibnamefont{Jackson}}, \bibnamefont{and}
  \bibinfo{author}{\bibfnamefont{S.}~\bibnamefont{Stringari}},
  \bibinfo{journal}{Phys. Rev. A} \textbf{\bibinfo{volume}{73}},
  \bibinfo{pages}{013603} (\bibinfo{year}{2006}),
  \href{http://www.arxiv.org/abs/cond-mat/0510143v1}{\nolinkurl{arXiv:cond-mat%
/0510143v1}}.

\bibitem[{\citenamefont{Nelson et~al.}(2007)\citenamefont{Nelson, Li, and
  Weiss}}]{NelLiWei07}
\bibinfo{author}{\bibfnamefont{K.~D.} \bibnamefont{Nelson}},
  \bibinfo{author}{\bibfnamefont{X.}~\bibnamefont{Li}}, \bibnamefont{and}
  \bibinfo{author}{\bibfnamefont{D.~S.} \bibnamefont{Weiss}},
  \bibinfo{journal}{Nat.~Phys.} \textbf{\bibinfo{volume}{3}},
  \bibinfo{pages}{556} (\bibinfo{year}{2007}).

\bibitem[{\citenamefont{{van Oosten} et~al.}(2001)\citenamefont{{van Oosten},
  {van der Straten}, and Stoof}}]{Oosten-PRA-2001}
\bibinfo{author}{\bibfnamefont{D.}~\bibnamefont{{van Oosten}}},
  \bibinfo{author}{\bibfnamefont{P.}~\bibnamefont{{van der Straten}}},
  \bibnamefont{and} \bibinfo{author}{\bibfnamefont{H.~T.~C.}
  \bibnamefont{Stoof}}, \bibinfo{journal}{Phys. Rev. A}
  \textbf{\bibinfo{volume}{63}}, \bibinfo{pages}{053601}
  (\bibinfo{year}{2001}),
  \href{http://www.arxiv.org/abs/cond-mat/0011108v1}{\nolinkurl{arXiv:cond-mat%
/0011108v1}}.

\end{thebibliography}

\end{document}